\documentclass[final,5p,times]{elsarticle}

\usepackage{amssymb}
\usepackage{amsmath}

\usepackage{array}

\usepackage{graphicx}
\usepackage{subcaption}

\usepackage{mathrsfs}
\usepackage{braket}
\DeclareMathOperator*{\argmin}{arg\,min}

\usepackage[hyphens]{url}
\usepackage{hyperref}
\hypersetup{
    colorlinks=false,
    allcolors=black,
    pdfborder={0 0 0}
}

\newcommand\blfootnote[1]{%
  \begingroup
  \renewcommand\thefootnote{}\footnote{#1}%
  \addtocounter{footnote}{-1}%
  \endgroup
}

\begin{document}
\begin{frontmatter}

\title{\vspace{-0.4em}Enhanced fill probability estimates in institutional algorithmic bond trading\\using statistical learning algorithms with quantum computers}

\cortext[author_cor]{Corresponding authors.}
\author[author_hsbc_uk1]{Axel Ciceri}
\author[author_hsbc_uk1]{Austin Cottrell}
\author[author_hsbc_uk1]{Joshua Freeland}
\author[author_ibm_romania]{Daniel Fry}
\author[author_ibm_japan]{Hirotoshi Hirai}
\author[author_hsbc_uk2]{Philip Intallura\corref{author_cor}}
\author[author_ibm_us]{Hwajung Kang}
\author[author_ibm_us]{\\Chee-Kong Lee}
\author[author_ibm_us]{Abhijit Mitra}
\author[author_ibm_japan]{Kentaro Ohno}
\author[author_ibm_us]{Das Pemmaraju}
\author[author_ibm_ch]{Manuel Proissl\corref{author_cor}}
\author[author_ibm_us]{Brian Quanz}
\author[author_hsbc_uk2]{Del Rajan}
\author[author_ibm_japan]{\\Noriaki Shimada}
\author[author_ibm_india]{Kavitha Yograj}

\affiliation[author_hsbc_uk1]{organization={HSBC Holdings Plc., Credit Algorithmic Trading}, country={United Kingdom}}
\affiliation[author_hsbc_uk2]{organization={HSBC Holdings Plc., Emerging Technology, Innovation, and Ventures}, country={United Kingdom}}
\affiliation[author_ibm_us]{organization={IBM Quantum, IBM Research}, country={United States of America}}
\affiliation[author_ibm_ch]{organization={IBM Quantum, IBM Research Europe}, country={Switzerland}}
\affiliation[author_ibm_romania]{organization={IBM Quantum, IBM Research}, country={Romania}}
\affiliation[author_ibm_japan]{organization={IBM Quantum, IBM Research}, country={Japan}}
\affiliation[author_ibm_india]{organization={IBM Quantum, IBM Research}, country={India}}

\begin{abstract}
The estimation of fill probabilities for trade orders represents a key ingredient in the optimization of algorithmic trading strategies. It is bound by the complex dynamics of financial markets with inherent uncertainties, and the limitations of models aiming to learn from multivariate financial time series that often exhibit stochastic properties with hidden temporal patterns. In this paper, we focus on algorithmic responses to trade inquiries in the corporate bond market and investigate fill probability estimation errors of common machine learning models when given real production-scale intraday trade event data, transformed by a quantum algorithm running on IBM Heron processors, as well as on noiseless quantum simulators for comparison. We introduce a framework to embed these quantum-generated data transforms as a decoupled offline component that can be selectively queried by models in low-latency institutional trade optimization settings. A trade execution backtesting method is employed to evaluate the fill prediction performance of these models in relation to their input data. We observe a relative gain of up to $\sim 34$\% in out-of-sample test scores for those models with access to quantum hardware-transformed data over those using the original trading data or transforms by noiseless quantum simulation. These empirical results suggest that the inherent noise in current quantum hardware contributes to this effect and motivates further studies. Our work demonstrates the emerging potential of quantum computing as a complementary explorative tool in quantitative finance and encourages applied industry research towards practical applications in trading.
\end{abstract}

\end{frontmatter}

\section{Introduction}
\label{sec_intro}
In the risk-conscious quantitative management of investment portfolios, we are often concerned with the chance of financial profits or losses from significant deviations of estimates about future market developments and the execution likelihood of trade orders that inform trading strategies. Thus, the dynamics of financial markets with associated fundamental uncertainties are subject to research with a long history and continuous advancements of mathematical models that aim to unravel the complexity of accessible market information and, under respective model uncertainties, to optimally contribute to deriving better trading decisions.\blfootnote{The authors are listed alphabetically.}\blfootnote{Contacts: \href{mailto:philip.intallura@hsbc.com}{\texttt{philip.intallura@hsbc.com}}, \href{mailto:manuel.proissl@ibm.com}{\texttt{manuel.proissl@ibm.com}}}

While it is widely acknowledged in financial economics that it is not possible to seek a universal theory that accurately describes the underlying process of time propagation in markets, most of the commercial success of quantitative investing and algorithmic trading suggests that it is still possible to find time- and scale-dependent local approximations using models derived from empirical observations with sufficient statistics. However, the bounding limitation resides both in the predictability of noisy financial observables with their higher-dimensional non-linear interactions that exhibit stochastic properties, and in the corresponding search for hidden temporal regularities in these interactions at varying scales and frequencies that could serve as trading signals.

The problem domain belongs to a long-lasting interdisciplinary quest that has prominently shown an intriguing phenomenological connection between financial and physical processes through advances in applied mathematics. It began with the seminal works on Brownian motion in the context of asset price fluctuations and option pricing~\cite{bachelier1900_theoryofspeculation}, and in the context of thermodynamics and atomic behavior~\cite{einstein1905_brownianmotion}, followed by partially independent progressions into stochastic analysis, probability theory, statistical learning, and quantum mechanics. The former three contribute to the cornerstone of modern quantitative finance and the modeling of financial observables, and, in the advent of artificial intelligence with advanced computing systems, are empowering innovations in trading today.

But until recently, at least beyond theoretically inspired explorations that would address the fundamentally bounding limitations noted above, any extension of modeling capabilities into the quantum realm\footnote{The \textit{quantum realm} can be thought of as a naturally inspired enhancement of probabilities with an enriched set of mathematical operations in an exponentially larger state space (see Section~\ref{subsec_quantumfeature}).} was inaccessible. Given the emergence of first commercially available quantum computers, although still limited by their inherent noise with respective mitigation~\cite{ibm2023_utilitypaper} but fast approaching fault tolerant scale~\cite{ibm2024_ftqm}, these emerging devices are starting to allow us to experimentally probe new ideas, inspire novel economic research directions and potentially enhance financial time series analysis by opening a gateway to a fundamentally different and widely unexplored regime of information processing and analysis of complex probability distributions underlying dynamical systems and processes.

This motivates the investigation presented in this paper, in which we use an actual quantum computer as an explorative tool following an empirical rather than theoretical approach. Thus, we only report statistical observations, aligned to known backtesting practices in the industry and a benchmarking setup that strictly isolates the quantum part from the classical part\footnote{The terms \textit{classical} and \textit{quantum} are used to distinguish the use of conventional and quantum computers or algorithms, respectively.} used in practice, and we do not infer any generalizable theory or causal economic effect that would explain these observations. As a consequence, our treatise does not expect the reader to be familiar with quantum mechanics or the inner workings of a quantum computer, even though we share the relevant self-contained theory and mathematical operations in Section~\ref{sec_method}. In fact, the analysis design and test setup introduced in Section~\ref{sec_setup} allow one to appreciate the results in Section~\ref{sec_results} in their own right, even if the quantum computer is considered as a hidden \textit{black box} function that can be queried reproducibly.

As our empirical analysis aims for a practically meaningful and realistic setting that deals with the complexity of market dynamics within a confined problem space, we focus on real and classically hard problem instances in algorithmic trading of European corporate bonds from a market-making perspective. In particular, the process of interest concerns a blind auction with trade inquiries sent over automated trading systems as RFQs (Request for Quote) and the subsequent algorithmic responses with a quote, conditional on the dealer's inventory interest and trading strategy. The automated pricing in this market is particularly challenging -- driven by institutional flows with insurance companies, asset managers, hedge funds, and banks -- and often deals with sparse pricing data and high dimensionality of inquiry characteristics and market conditions. In such a blind auction, every trader, automated or traditional, must weigh the desire to win business with the desire to profit from the business they execute. A key problem in this business is the \textit{execution likelihood estimation} or \textit{fill probability} of a given RFQ response. It represents a core component of a trading strategy and shapes the processing flow of a trading system, where any improvements in execution fidelity present a meaningful opportunity for competitive advantage, and would yield improved margins, improved risk management via a higher hit rate on preferred trades, avoidance of undesired transactions, and increased future deal flow due to improved competitiveness.

\begin{figure}[t!]
\centering
\includegraphics[width=\columnwidth]{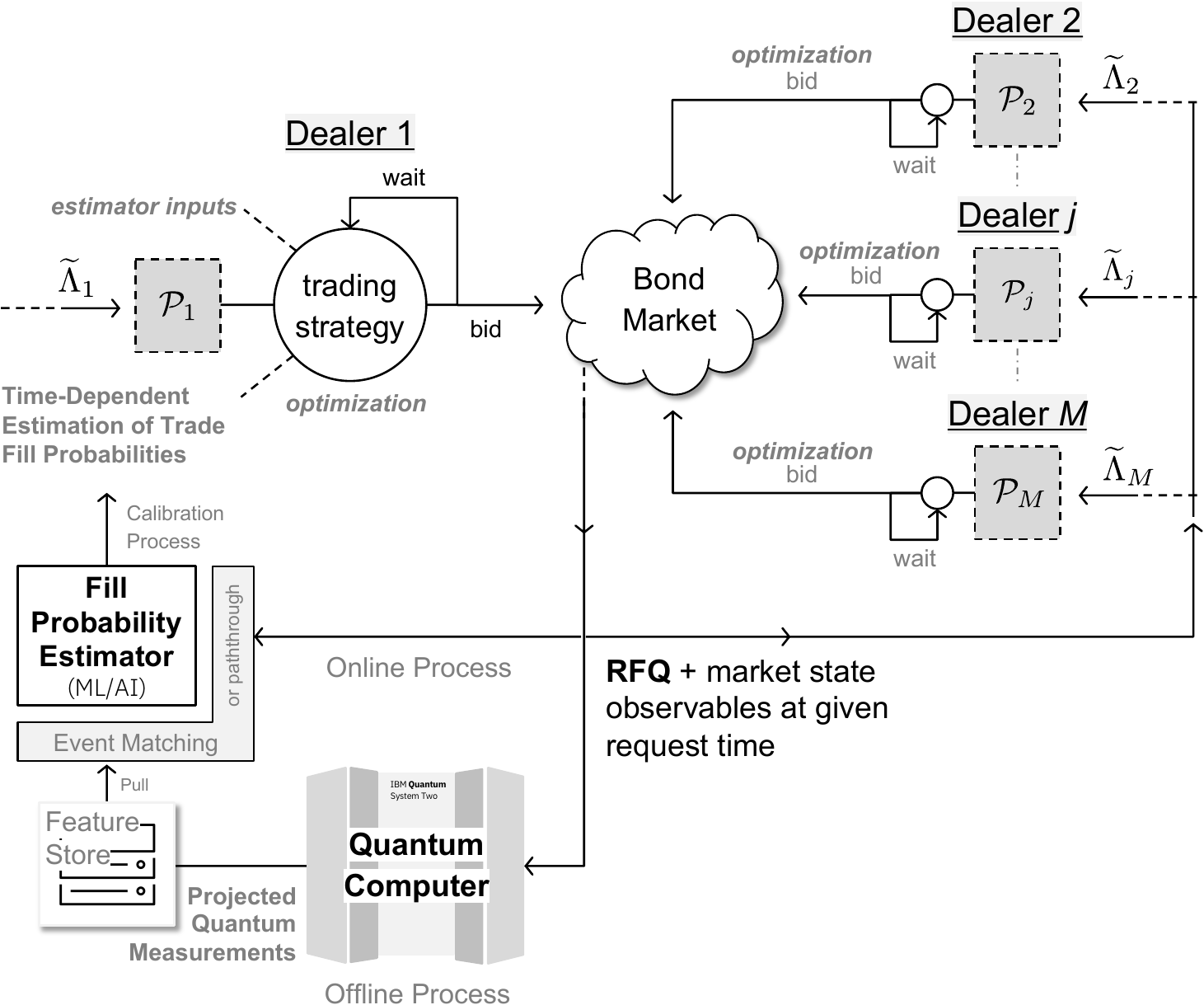}
\caption{Illustration of the business problem embedding explored in this work. Each market-making dealer responds to RFQs and influences their response and trading strategy through the individual fill probability estimator $\tilde{\Lambda}_i$. One of the dealers passes additionally RFQ characteristics and market information through a quantum computer that generates complementary features for subsequent use in $\tilde{\Lambda}_1$ with resulting $\mathcal{P}_1$ outputs. The left-hand side of this illustration represents the simulation performed by the backtesting protocol presented in Section~\ref{subsec_backtesting}.}\label{fig_bigpicture}
\end{figure}
But any statistical model to help estimate fill probabilities is fundamentally limited by the intrinsic complexity and limitations of corporate bond trading datasets. Since increasing the complexity of models often does not translate into better results for this problem, the data input to the models becomes subject to further investigation and raises the question: Can non-trivial data transformations help models to reduce errors in fill probability estimation? While we may not be able to provide an answer to this question, it inspires the idea for the exploration presented in this paper. As illustrated in Figure~\ref{fig_bigpicture}, we take trade event vectors that combine RFQ and current market information, pass them through a particular quantum algorithm executed on a quantum computer to generate a corresponding set of transformed trade event vectors, then use those instead to train a machine learning model for estimating fill probabilities, and finally, compare with actually realized outcomes. Our study uses a real intraday algorithmic bond trading dataset at full production scale and applies a backtesting approach commonly used in the industry to compare models with and without quantum-transformed datasets.

\section{Methodology}
\label{sec_method}
The method to enhance estimates of trade fill probabilities in electronic bond markets, as adopted in this paper, is presented in this section. Before formally defining the problem and heuristic solution approach, we motivate the incentives behind the chosen approach.

\subsection{Background and motivation}
\label{subsec_motivation}
Any prediction attempt of trade execution likelihoods deals with at least two key limiting factors: time and market state representation uncertainty. 

The first factor touches a major challenge of statistically learning and inferring the time evolution of financial observables given a selected batch of historical data. This comes from the fact that we accept the financial system, and any subset forming a market, to be irreversible in time\footnote{The argument can easily be followed when considering the definition of a financial market used in this paper: a complex dynamic interaction network of probabilistic human players or respective authors of trading algorithms that each evolve and adapt to changes of others over time, and randomly associate to different roles on buy- or sell-side, trade incentives, risk limits, long memory effects, and irrational behavior inducing variable responses to private and public information.} or not statistically invariant upon time reversal as empirical financial time series analyses suggest~\cite{zumbach2009_timereversal}. More formally, this means that past decisions, such as trade orders, do not generally yield the same outcome, such as associated order fills, at a later time, even if the observable market information appears to be the same. In other words, the true state of the past market never returns exactly and thus fundamentally bounds the predictability of future states and respective outcomes. But we can still assume, and test as in Section~\ref{sec_results}, that instances of positive time-dependent predictability exist, in which the time difference $\delta$ between learning from market information up to time $t$ and inferring outcomes given market information at $\tau > t$ influences the predictability negatively as $\delta$ grows or vice versa. Hence, the prediction error may scale probabilistically at some unknown rate as a function of time.

The second factor relates to the fundamental question of how to represent, or rather approximate, the state of a market in the first place, which would then seed time propagation in the modeling task and dominate the lower bound of the prediction error when considering an infinitesimal $\delta$. This naturally follows from the fact that, at any given time, a market participant has only access to data with partial market information from its vantage point and, due to algorithmic and computational constraints, needs to make a practically feasible hypothesis of selecting a subset of data features that would sufficiently describe a market state within the correspondingly spanned feature space, which then serves as some form of proxy for the unknown true market state (see Section~\ref{subsec_staterep}). But regardless of any subset choices, there is an unknown amount of missing information that may weakly or more strongly interact with the selected feature space at different times, and thus creates an uncertainty that is hard to measure or estimate. Another source of uncertainty comes from the time it takes to physically create a market state representation and to act upon its observation, during which the market state has evolved and collapsed into an unknown state at the time of response, such as executing a trade, for instance. Even if there exists a complex trading model that could account for such state drifts and a respective computing device that could encode an exponentially large feature space spanned by all publicly observable information, such as news events, empirical research still suggests~\cite{kyle1985_priceimpact, maitrier2025_priceimpact} that there are significantly many instances when market impact and price jumps appear without any association to prior events.

In this work, we are considering these foundational limitations from a market maker perspective in the context of an electronic auction of corporate bonds with the following practical challenges: a hidden distribution of trade inquiries to selected market participants, inquiry response time windows at the order of seconds, lack of response price transparency among all competing bidders, and limited information about the acceptance or rejection of submitted offers. This creates a sparse and statistically irregular landscape of time-dependent information for models to learn from and motivates our approach to fully decouple the representation of a market state from learning a function that associates a given state with an offer acceptance likelihood or fill probability, based on observing similar near-past states and respective auction response outcomes.

Such a decoupling enables then the creation of a real-time response system, as anticipated in Figure~\ref{fig_bigpicture}, with an offline component that generates discriminative features to represent market states, and then gets queried by an online component that either infers incoming inquiries with a learned fill probability estimator function $\tilde{\Lambda}$ or recalibrates it to adapt to new market state-to-outcome associations. Since the composition of these features is based on the hypothesis that they jointly encode relevant market signals for the estimation task, then the prime limiting factor becomes the search for a function that finds those hidden signals to discriminate states that are more likely to yield a trade response to be filled. Therefore, the motivation and central piece of our approach is devoted to the offline component that uses a quantum algorithm to generate transformed features independently of the outcome associations, which aim to expose market signals such that the complexity of searching for them is reduced and thus would create an enlarged space of learnable fill probability functions with a lower prediction error than those without access to these quantum-generated features.

Our approach also addresses practical challenges in the industry, such as model complexity considerations in model risk management and regulatory controls. This is because the envisioned quantum enhancement is targeting only updates to the input data for a model to learn from and not necessarily the modeling methodology itself, meaning no changes to the currently used learning algorithm would be a priori required, and a potentially less complex or more robust model may be chosen, which keeps the model complexity manageable.

\subsection{Problem formulation}
\label{subsec_problem}
\begin{figure}[t!]
\centering
\includegraphics[width=\columnwidth]{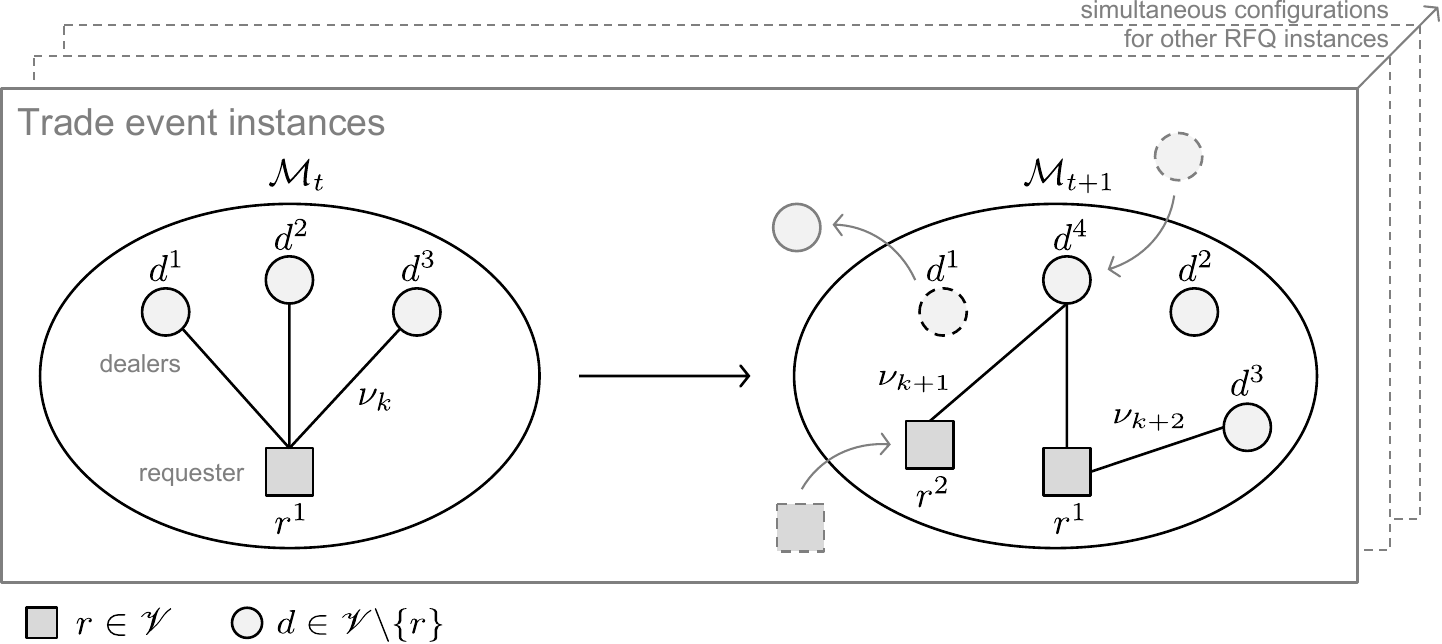}
\caption{Conceptional view of the dynamics of $\mathcal{M}_t$ and associated interactions between RFQ issuers or requesters $r$ and dealers $d$ over time. The superscripts are used as labels to uniquely distinguish the players in this view.}\label{fig_trade_instance_dynamics}
\end{figure}

Consider an electronic bond market as a graph $\mathcal{M}_t = (\mathscr{V}, \mathscr{E})_t$ propagating in continuous time $t$ with an open community of market participants $\mathscr{V}_t = \{1, \dots, N_t \} \in \mathbb{Z}^{+}$ of time-dependent size $|\mathscr{V}_t| > 1$ and temporal interactions $\mathscr{E}_t \subseteq \mathscr{V}^2_t$, with $a \neq b$ $\forall \{ a,b \} \in \mathscr{E}_t$, that get instantiated by a unique RFQ process identified by $k \in \mathbb{N}$ at time $t_k$, where a requesting entity $r_k \in \mathscr{V}_t$ opens a blind auction with selected dealers $D_k \subset \mathscr{V}_t \backslash \{r_k\}$ that synchronously receive a trade inquiry in the form of an electronic information blob or vector $\nu_k$ that maps to specific RFQ configurations. These include the globally unique bond identifier, desired trade size, buy- or sell-side direction, time allotted for response by $D_k$, time allotted for decision making by $r_k$, settlement details, and specified quoting convention, such as price, yield, or yield spread to a reference benchmark. Each dealer $d \in D_k$ then algorithmically decides to either respond with a quote $q^d_k$ or not, which generates a response set $Q_k \subset \{ q^1_k, \dots, q^{|D_k|}_k \} \cup \emptyset$, on which basis $r_k$ at $t^\prime_k > t_k$ accepts an offer to trade from $Q_k$, in case it is non-empty, or leaves its RFQ $k$ expire untraded. This closes the auction and the respective participating relations associated with $k$ in $\mathscr{E}_{t^\prime}$ vanish. This conceptual view, in which we embed our problem of interest, is illustrated in Figure~\ref{fig_trade_instance_dynamics}.

We make no assumptions about the dynamics of $\mathcal{M}_t \subset \mathcal{S}_t$, its interactions with other markets of the financial system $\mathcal{S}_t$, and any global structure that would bound the space of collective trading strategies or algorithms seeking Nash equilibria. Instead we construct the problem on a purely observational information-centric and probabilistic basis using heuristic methods, and do so only from the local perspective of a single dealer $d \in \bigcup D_k$ for all $k \in \{k_i, \dots, k_j\} =: K_\mathfrak{T}$ that $d$ gets selected during a given trading time period $\mathfrak{T} := [0, T]$ of $(\mathcal{M}_t)_{t \in \mathfrak{T}}$ with $T > 0$. For each incoming RFQ $k$ with associated $\nu_k$ that may be of interest to the current inventory of bonds, the dealer aims to optimize some form of a utility response function $U(\nu_k, \cdot)$ as part of an algorithmic trading strategy $(\xi_t)_{t \in \mathfrak{T}}$ that intends to balance risk and profitability objectives under temporal market conditions and business constraints. As a key ingredient in this optimization problem, let $\tilde{\Lambda}$ be a fill probability estimator that assigns acceptance likelihoods in $[0,1]$ to permissible quote candidates $Q^d_k$, such as a bounded response price range in $\mathbb{R}$. The estimator uses as inputs market state representations $X$ that encode information from $\nu_k$ and respective market conditions, as described in Section~\ref{subsec_staterep}, and is trained with past realized outcomes $y \in \{0, 1\}$ that label accepted RFQ responses with a "1" and otherwise "0", as described in Section~\ref{subsec_tradelearning}.

\begin{figure}[t!]
\centering
\includegraphics[width=\columnwidth]{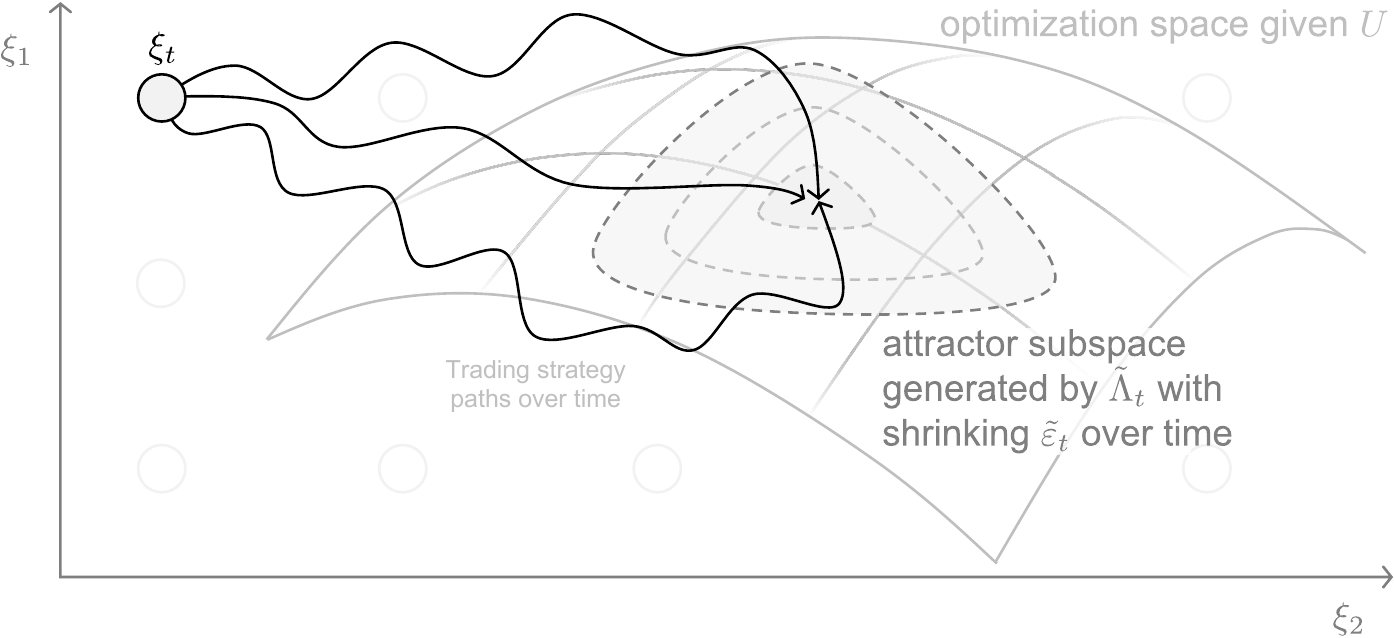}
\caption{Business objective behind fill probability estimation: given a trading strategy $\xi_t$ over time period $\mathfrak{T}$ and its optimization space spanned by $U$, attract or bias the paths of $\xi_t$ by generating smaller optimization attractor subspaces (depicted here by contours) with shrinking $\tilde{\varepsilon}_t$ of $\tilde{\Lambda}_t$ estimators that learn iteratively from realized RFQ response outcomes over time $t$. The term \textit{attractor subspace} is used to imply a dynamic field that locally attracts but not globally constrains the time evolution of optimization paths towards the terminal state objectives of $\xi_t$.}
\label{fig_trading_strategy_fillprob_errors}
\end{figure}

Then $\tilde{\Lambda}$ acts as a filtration of potential responses and confines the optimization space or loss landscape of $U$ to a more relevant $\xi_t$-path-attracting subspace that, conditional on the estimated prediction error $\tilde{\varepsilon} := \mathbb{E}[|\tilde{\Lambda} - \Lambda|]$ with respect to the realized truth $\Lambda$, induces over time a trade preference bias in $\xi_t$ such that it tends to steer more often towards deals with a higher chance of winning and conversely, a lower chance of losing deals that leak information about the dealer's trade interests into the market. This objective is illustrated in Figure~\ref{fig_trading_strategy_fillprob_errors}.

However, we are not concerned about the nature of $\xi_t$ in our study, but about the reduction of $\tilde{\varepsilon}$ with respect to transformations $\phi$ of $X$, which then may indirectly influence the feasible paths that a particular $\xi_t$ could take. Essentially, one seeks to $\text{min}_{\phi}\tilde{\varepsilon}$ subject to sets of $\phi (X)$, of which one is the identity function $\phi_0 (X) = X$ with $\tilde{\varepsilon}_0$ to benchmark against. This offers the problem for experimental tests in this paper: find a transformation $\phi_q$ using a quantum algorithm that yields $\tilde{\varepsilon}_q < \tilde{\varepsilon}_0$ for given $X$ over a confined trading period of bond RFQs.

\subsection{Market state representations}
\label{subsec_staterep}
There is no prescription to represent market states -- only a hypothesis about information that seems relevant to explain the process of interest can be made. We are interested in the auction process described before, which already provides a vector $\nu_k$ of information for a given RFQ $k$ at incoming time $t_k$. This vector is considered to already only contain characteristics about the RFQ, rather than $k$-specifics, such as the ISIN (International Securities Identification Number) of the particular bond. Providing a $\tilde{\Lambda}$ estimator only with this information, the problem would likely converge to a game of random guessing, since all other receiving dealers have the same information. Therefore, we use $\nu_k$ along with the ISIN to map additional features at the given time of the market, and thus extend this vector into, thereafter referred to, a market state representation or \textit{event} $x \in \mathbb{R}^p$ with $p$ number of elements or features.

Each feature of an event is considered a random variable, propagating in time. In order to detect structural shifts and asymmetries in a collection of past events over time, the composition of features is based on the extraction of time series characteristics from the time $t_k$ backwards. This includes scans of time-lagged macroeconomic and financial observables, trends, and correlations between near- to far time blocks of observables such as price movements and shifts in entropy, or changes in sequences of aggregations such as past buy- or sell-side RFQ response statistics. In fact, any information could be selected, even past outcomes $y_t$ with $t < t_k$, since it is itself a random variable.

More formally, we are considering a well-defined filtered probability space $(\Omega, \mathcal{F}, (\mathcal{F}_t)_{t \in \mathfrak{T}}, P)$ with a finite sample space $\Omega$ of possible market states, a $\sigma$-algebra $\mathcal{F}$ as the measurable event space with a filtration to our trading window, and a probability measure $P$. We refer to $X \in \mathbb{R}^{s \times p}$ with $s$ number of events as an additional filtration of the event space that further discretizes the spread of observations in $\mathfrak{T}$ we can learn from. This means that not all received RFQs $K_{\mathfrak{T}}$ are visible for training $\tilde{\Lambda}$. Therefore, to approximate maps from events in $X$ to associated outcomes $y$, the learner must work with limited local statistics before $t_k$ and needs to leverage the constructed event features, as described above, which contain continuous time information.

\subsection{Trade execution learning}
\label{subsec_tradelearning}
On the basis of a given trading dataset, containing the event matrix $X$ that is paired with the event outcome vector $Y$ of binary values $y \in \{0,1\}$, labeling filled trade instances with "1" or unfilled with "0", we describe in this section the construction of the fill probability estimator $\tilde{\Lambda}$ using a machine learning approach~\cite{hastie2009elements} in more detail.

First, we note that the fill probability in this setup is governed by some unknown, underlying joint probability distribution $P(X, Y)$ that can change with time, and likewise, $P(Y | X)$ can change at some unknown rate, even during the same market regime. The learning setup, therefore, relies on the rich structure of the feature space spanned by $x \in \mathbb{R}^p$ to distinguish differences in time and respective market conditions. Since there is only partially observable information in $x$ and underlying uncertainties, the objective of the learning task is to find a suitable approximation of $P(Y | X)$ within a rolling time window of the market, such that outcome labels of future unseen events can be predicted as accurately as possible. As further described in Section~\ref{subsec_backtesting}, a particular backtesting protocol is used for such rolling in time.

More formally, let $\mathcal{Z}^s$ be a space of event-outcome tuples, combining $X$ and $Y$, and $\mathcal{D} \in \mathcal{Z}^s$ a dataset sample with $s$ number of state pairs $\{ (x_i, y_i) \}_{i=1}^s$. The objective is then to learn a function $\tilde{\Lambda}: \mathbb{R}^p \rightarrow [0, 1]$ that maps an event $x \in \mathbb{R}^p$ to a corresponding fill probability $P(y=1|x) \in [0,1]$. In order to estimate $\tilde{\Lambda}$, a particular parametric model is assumed, such as a neural network or a decision tree, with one or more model parameters $\theta$. The model choice, which is explained later in Section~\ref{subsec_backtesting}, defines a hypothesis space $\mathcal{H}$. Given this choice, the actual learning procedure is then encoded in an algorithm $\mathcal{A} : \mathcal{Z}^s \rightarrow \mathcal{H}$ that maps samples $\mathcal{D} \in \mathcal{Z}^s$ to functions $\tilde{\Lambda}_{\theta} \in \mathcal{H}$. Concretely, the algorithm $\mathcal{A}$ involves fitting $\tilde{\Lambda}_{\theta}$ to a training sample $\mathcal{D}$ by optimizing the model parameters $\theta$ in order to minimize a loss function, $\ell(y, \tilde{\Lambda}_{\theta}(x))$ averaged over the training sample, i.e., $\argmin_\theta \sum_i \ell(y_i, \tilde{\Lambda}_{\theta}(x_i))$.  For example, the logistic loss function,
\begin{equation*}
\ell(y, \tilde{\Lambda}_{\theta}(x)) = -\left( y \log(\tilde{\Lambda}_{\theta}(x)) + (1 - y) \log(1 - \tilde{\Lambda}_{\theta}(x)) \right),
\end{equation*}
is a common choice, which corresponds to maximizing the log-likelihood of observing the data $\mathcal{D}$ under the given model, $\tilde{\Lambda}_{\theta}$.

While the learning process involves minimizing the empirical loss over the training data in $\mathcal{D}$, the goal is to arrive at an estimator $\tilde{\Lambda}_{\theta}$ that can generalize, meaning to remain accurate on unseen test data $(x_{\text{test}}, y_{\text{test}}) \sim P(X,Y)$, as quantified by a low expected test loss $\mathbb{E}_{P(X,Y)}[\ell(y_{test}, \tilde{\Lambda}_{\theta}(x_{\text{test}}))]$.

To this end, one needs to ensure that the overall learning framework $\{\mathcal{H}, \mathcal{A}\}$ adequately manages the \textit{bias-variance} tradeoff -- namely the inherent trade-off between adopting a large hypothesis space $\mathcal{H}$ and obtaining an estimator $\tilde{\Lambda}_{\theta}$ with a strong dependence on the finite training set $\mathcal{D}$. In this work, we manage the bias-variance tradeoff through regularization techniques that limit the expressivity of $\mathcal{H}$ while augmenting $\mathcal{A}$ through cross-validation. For instance, for logistic regression, we incorporate $L^2$ regularization via the loss function $\ell$, while in tree-based models, we limit the tree-depth.

At the end of this procedure, the obtained hyperparameter optimized estimator $\tilde{\Lambda}_{\theta}^{\text{opt}}$ (or referred to only as $\tilde{\Lambda}$ in the paper) is used to evaluate its prediction performance on unseen test data. Standard learning theory offers generalization guarantees with respect to making in-distribution predictions, i.e., under the assumption that both observed training and unseen test samples are i.i.d. sampled from the same distribution $P(X,Y)$, and thus bounding the error $\mathbb{E}_{P(X,Y)}[\ell(\tilde{\Lambda}_{\theta}(X),Y)]$. However, as discussed before, this is not applicable in our problem domain as we cannot assume to have i.i.d. access to the full underlying distribution $P$ at training time. It may well be the case that the set of training inputs $X_{\text{train}}$ from observed historical market states are sampled from an apparent distribution $P_{\text{history}}$ with restricted support as in $\mathrm{supp}_{X_{\text{train}}}(P_{\text{history}}) \subsetneq \mathrm{supp}_{X}(P)$. In such a scenario, the generalization bounds of learning theory only apply to test data also drawn from $P_{\text{history}}$ and we can only bound $\mathbb{E}_{P_{\text{history}}(X,Y)}[\ell(\tilde{\Lambda}_{\theta}(X),Y)]$ but not $\mathbb{E}_{P(X,Y)}[\ell(\tilde{\Lambda}_{\theta}(X),Y)]$.

But the business problem at hand only cares about fill probability estimates in financial time series with a strict decoupling of train and test sets by market time. This is to remind the reader that this time propagation makes the elements of $x$ and $y$ random variables in the first place and are subject to the specific nature of underlying stochastic processes discussed in Section~\ref{subsec_motivation} and \ref{subsec_problem}. As a direct consequence, \textit{no general and data-instance independent theoretical guarantees} on prediction errors $\tilde{\varepsilon}_0$ can be made for such out-of-sample or out-of-distribution tests, where respective models may incorporate temporal domain knowledge in the form of a suitable inductive bias. Therefore, we are adopting a \textit{purely empirical approach} and design our experimental tests using backtesting, commonly used in the financial industry.

\subsection{Quantum feature generation}
\label{subsec_quantumfeature}
Central to our trading data-specific exploration of a \textit{quantum enhanced} approach is the notion of a feature transformation $\phi:\mathbb{R}^p \rightarrow \mathbb{R}^q$ enabled by a quantum device. Such a feature transformation enters via a decomposition of the desired mapping $x \xrightarrow{\tilde{\Lambda}} P(y=1|x)$ defined above, into a two step process $x \xrightarrow{\phi} x' \xrightarrow{g} P(y=1|x')$, i.e., $\tilde{\Lambda} \coloneqq g \circ \phi$ with $\phi$ assumed deterministic for simplicity. In other words, the mapping $x \xrightarrow{\phi} x'$ represents an explicit \textit{feature engineering} step, which is beneficial if we assume that, from the perspective of the learning algorithm $\mathcal{A}$ and hypothesis class $\mathcal{H}$ at hand, the mapping $x' \xrightarrow{g} P(y=1|x')$ on dataset $\mathcal{D}' = \{(x'_i, y_i)\}_{i=1}^s$ is in some sense easier to estimate than the original $\tilde{\Lambda}$. In this setting, quantum feature generation then deals with the internal structure of the map $\phi: \mathbb{R}^p \rightarrow \mathbb{R}^q$. Before we dwell into how $\phi$ is constructed in the next paragraph, we emphasize that both the original input features $x \in \mathbb{R}^p$ and the transformed features $x'\in\mathbb{R}^q$ are classical quantities that can be stored and accessed in classical memory. So, for understanding the empirical results of this work, as it relates to the fill probability estimation problem, $\phi$ can be safely assumed to be an arbitrary blackbox transformation. 

In order to describe the internal structure of $\phi$, we first start with a set of relevant vector spaces:
\begin{itemize}
    \item the original input feature vectors $x$ are drawn from a data space $\mathcal{X} \subset \mathbb{R}^p$, corresponding to vectors with $p$ number of features;

    \item on the quantum computer, we have access to a Hilbert space\footnote{A Hilbert space is a complete inner product space -- generalizing concepts of Euclidean geometry in 2D or 3D to spaces with any number of dimensions~\cite{debnath2005introduction}.} $\mathcal{H}_Q$, whose elements are quantum states $|\psi\rangle \in \mathcal{H}_Q$, and also access to operators $\hat{O}$ from $\mathcal{B}(\mathcal{H}_Q)$, the space of bounded linear operators on $\mathcal{H}_Q$;

    \item the transformed features $x'$ live in an image space $\mathcal{X}' \subset \mathbb{R}^q$, corresponding to vectors with $q$ number of features.
\end{itemize}
Note that the quantum state is represented by a complex vector, and herein we use common notation in quantum mechanics, where $|\psi\rangle$ indicates a vector and $\langle \psi |$ its Hermitian conjugate, so that, for example, the inner product of $|\varphi\rangle$ and $A |\psi\rangle$ for an operator $A$ can be succinctly expressed as $\langle \varphi | A | \psi \rangle$. The quantum Hilbert space $\mathcal{H}_Q$ has a dimensionality that is exponential in the number of qubits $N_Q$, the basic units of information on the quantum device, which means that in general quantum states $|\psi\rangle \in \mathcal{H}_Q$ cannot be efficiently stored in classical memory. We note that within the quantum framework, the states $|\psi\rangle \in \mathcal{H}_Q$ encode quantum probability distributions.  For reference, further details on quantum states, quantum computing, and the underlying mathematics can be found in~\cite{nielsen2010quantum}.

Given this setting, we choose the transformation $\phi$ to be a \textit{Projected Quantum Feature Map} (PQFM), which is a composite map 
\begin{equation}\label{eq_pqfm}
\phi \coloneqq \phi_{\mathcal{M}} \circ \phi_{\mathcal{U}} : \mathcal{X} \xrightarrow{\phi_{\mathcal{U}}} \mathcal{H}_Q \xrightarrow{\phi_{\mathcal{M}}} \mathcal{X}',
\end{equation}
consisting of an initial quantum embedding $|\psi\rangle = \phi_{\mathcal{U}}(x)$ that maps feature vectors $x$ to high-dimensional quantum states $|\psi\rangle$, followed by a projective measurement map $x' = \phi_{\mathcal{M}}(|\psi\rangle)$ that takes states $|\psi\rangle$ to transformed features $x'$. PQFMs are related to the notion of \textit{Projected Quantum Kernels}, first introduced in~\cite{huang2021_powerofdata}.

On an ideal quantum device, the process of mapping input vectors $x$ to quantum states is achieved by executing a quantum linear map $\hat{U}_{x, \alpha} \in \mathcal{B}(\mathcal{H}_Q): \mathcal{H}_Q \rightarrow \mathcal{H}_Q$, which takes quantum states to other quantum states unitarily. In particular, the quantum device is initialized in some fixed \textit{fiducial state} $|\psi_0\rangle \in \mathcal{H}_Q$ and the unitary $\hat{U}_{x, \alpha}$ -- parameterized by the inputs $x \in \mathbb{R}^p$ and optionally by additional hyperparameters $\alpha$ -- is executed to \textit{evolve} $|\psi_0\rangle$ into $|\psi_{x, \alpha}\rangle \in \mathcal{H}_Q$, i.e., 
\begin{equation*}
|\psi_{x, \alpha}\rangle=\hat{U}_{x, \alpha}|\psi_0\rangle = \phi_{\mathcal{U_{.,\alpha}}}(x).
\end{equation*}
This completes the data embedding step with the prepared state $|\psi_{x, \alpha}\rangle$ now parameterized by the inputs $x$. Next, in order to specify $\phi_{\mathcal{M}}$, we describe a quantum measurement protocol $\mathcal{M}$ that is characterized by a fixed set of $q$ \textit{operator observables} $\{\hat{O}_j\}_{j=1}^q$, where $\hat{O}_j \in \mathcal{B}(\mathcal{H}_Q)$. Operators $\hat{O}_j$ correspond to quantum random variables whose spectrum can be accessed by quantum measurements. Each component $x'_j$ of the transformed feature vector $x' \in \mathcal{X}'$ is then defined to be the expectation value of one of the operators $\hat{O}_j$ with respect to a state $|\psi\rangle \in \mathcal{H}_Q$, i.e., $x'_j \coloneqq \mathbb{E}[\hat{O}_j]_{|\psi\rangle} \coloneqq \langle \psi | \hat{O}_j |\psi\rangle$. Such expectations are obtained in practice by repeatedly preparing the state $|\psi\rangle$ and measuring the operator $\hat{O}_j$, while averaging over the realized outcomes from its spectrum (where repetitions are referred to as \textit{shots}). Thus, we have 
\begin{equation*}
    x' = \{x'_1, x'_2,...,x'_q\} = \{\mathbb{E}[O_1], \mathbb{E}[O_2],...,\mathbb{E}[O_q]\}_{|\psi\rangle} = \phi_\mathcal{M}(|\psi\rangle).
\end{equation*}
The complete PQFM $\phi$ is obtained when the state $|\psi_{x, \alpha}\rangle$, parameterized by the input features $x$, is substituted for $|\psi\rangle$ i.e.,
\begin{equation*}
x' = \phi_{\mathcal{M}}(|\psi_{x, \alpha}\rangle) = \phi_{\mathcal{M}}(\hat{U}_{x, \alpha}|\psi_0\rangle)=\phi_{\mathcal{M}}(\phi_{\mathcal{U_{.,\alpha}}}(x)).
\end{equation*}

In order to concretely implement a PQFM, we need to make specific choices for the embedding unitary $\hat{U}_{x, \alpha}$, the fiducial state $|\psi_0\rangle$, and the set of measurement observables $\{\hat{O}_i\}_{i=1}^q$ that are compatible with an actual quantum circuit architecture executable on the quantum device. Therefore, to implement $\hat{U}_{x, \alpha}$, we employ the parameterized circuit associated with a so-called \textit{Heisenberg ansatz} (HA). It is inspired by the one-dimensional Heisenberg model~\cite{heisenberg1928}, whose energy for a system with $N$ qubits is specified by the Hamiltonian operator
\begin{equation*}
    \hat{H}_{J} = \frac{1}{4}\sum_{j=1}^{N-1} J_j~\mathbf{\hat{S}}_j.\mathbf{\hat{S}}_{j+1} = \frac{1}{4}\sum_{j=1}^{N-1} J_j~\lbrack\hat{X}_j\hat{X}_{j+1} + \hat{Y}_j\hat{Y}_{j+1} + \hat{Z}_j\hat{Z}_{j+1}\rbrack,
\end{equation*} 
where the operators $\mathbf{\hat{S}}_j = \{\hat{X}_j,\hat{Y}_j,\hat{Z}_j\}$ are vectors composed of Pauli operators on qubit $j$, and the couplings $J_j$ are real parameters collectively represented by the vector $J$. The unitary operator corresponding to the HA is obtained via complex exponentiating the Heisenberg Hamiltonian, formally
\begin{align*}
    \hat{U}_{J, \alpha} &= \exp(-i \alpha \hat{H}_{J}) \\
    & = \exp\left(-i\frac{\alpha}{4}\sum_{j=1}^{N-1} J_j~\lbrack\hat{X}_j\hat{X}_{j+1} + \hat{Y}_j\hat{Y}_{j+1} + \hat{Z}_j\hat{Z}_{j+1}]\right) \\
    & = \exp\left(-i\frac{\alpha}{4}\left[\sum_{j=\text{odd}}^{N-1} \hat{H}_j(J_j) + \sum_{j=\text{even}}^{N-1} \hat{H}_j(J_j)\right]\right),
\end{align*}
where $\alpha \in \mathbb{R}^{+}$, and in the last line, the summations over odd and even site indices are grouped together with $\hat{H}_j(J_j) = J_j~\lbrack\hat{X}_j\hat{X}_{j+1} + \hat{Y}_j\hat{Y}_{j+1} + \hat{Z}_j\hat{Z}_{j+1}\rbrack$. Due to the non-commutativity of quantum operators, in general, the exponential of a sum of operators is not a product of exponentials involving the individual terms in the sum, but via Trotterization~\cite{suzuki1976}, we arrive at an approximate product formula for the unitary 
\begin{equation*}
    \hat{U}_{J, \alpha} = \Bigg( \prod\limits_{j=\text{odd}}^{N-1} \hat{U}_j(J_j) \prod\limits_{j=\text{even}}^{N-1} \hat{U}_j(J_j)\Bigg)^M
\end{equation*}
with $\hat{U}_j(J_j) = \exp(-i\frac{\alpha}{4M}\hat{H}_j(J_j))$, which is valid when $M \in \mathbb{N}$ is sufficiently large or $\alpha$ is sufficiently small. This unitary contains a number of repetitions $M$ of a basic unit enclosed within the large parentheses. For use within a PQFM, however, only the general product form of the unitary $\hat{U}_{J, \alpha}$ is important, and magnitudes of $M$ and $\alpha$ are not relevant. By identifying the vector of couplings $J$ of the HA, with individual input feature vectors $x$ from our dataset $\mathcal{D}$, we immediately arrive at a concrete form for the PQFM embedding unitary $\hat{U}_{x,\alpha}$ that we need. In making this identification, we further note that for our purposes, there is no need to strictly confirm to the physical symmetries of the original Heisenberg model. This means that the scalar $J_j$ at each site $j$ could also be replaced by a 3-tuple $(J_{j}^x, J_{j}^y, J_{j}^z)$ of real parameters yielding inhomogeneous pairwise couplings $\hat{H}_j(J_j) = \lbrack J_{j}^x\hat{X}_j\hat{X}_{j+1} + J_{j}^y\hat{Y}_j\hat{Y}_{j+1} + J_{j}^z\hat{Z}_j\hat{Z}_{j+1}\rbrack$ at each site $j$. The overall coupling vector $J$ could then be understood as a flattened grouping of such 3-tuples across all the qubits, which allows for a more qubit-efficient encoding of input features $x$ into the circuit ansatz.

\begin{figure}[t!]
\centering
\includegraphics[width=\columnwidth]{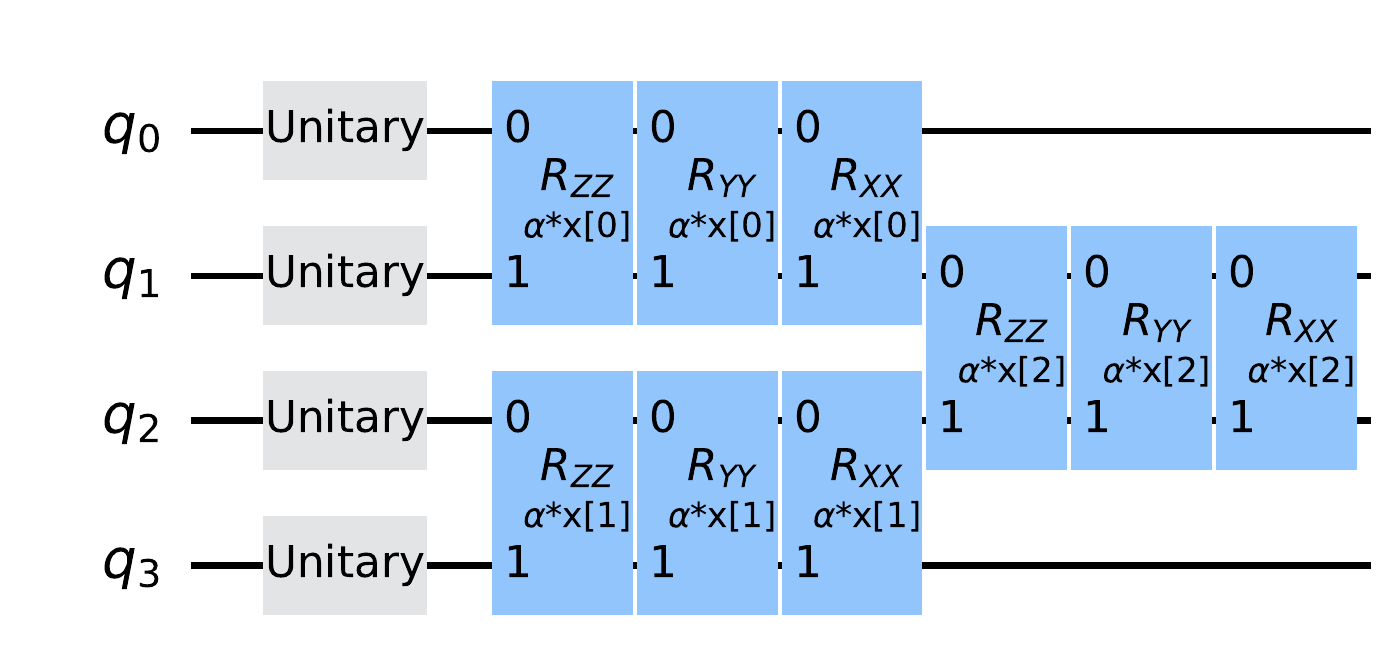}
\caption{Example of the quantum feature map circuit, following our \textit{Heisenberg ansatz} (HA), for 4 qubits and 3 features. Here, "Unitary" represents random single-qubit unitaries to set the initial state of each qubit (fiducial state), which can be tuned for a particular task. The sequence of $R_{ZZ}$, $R_{YY}$, $R_{XX}$ operations correspond to $\hat{U}_j$ for the indicted pair of qubits ($j$ and $j+1$) and the $i^{th}$ feature $x[i]$ (pattern/sequence of assigning features to each operation is arbitrarily chosen). These can be thought of as entangling two-qubit rotations about $ZZ$, $YY$, and $XX$, with the feature value controlling the rotation angle. Finally, $\alpha$ is a single shared additional parameter for the encoding that scales all rotations.}
\label{fig_heis_ex}
\end{figure}

In this work, we typically employ one or only a few repetitions $M$ and distribute the available feature vector components $x_k$ both over qubit indices and repetition indices. As the fiducial state, we set the tensor product state as
\begin{equation*}
|\psi_0\rangle = \bigotimes\limits_{j=1}^N |\psi_j\rangle,
\end{equation*}   
where $|\psi_j\rangle$ is a fixed but uniformly random single qubit state on qubit $j$. Through the fiducial state $|\psi_0\rangle$ and the HA unitary $\hat{U}_{x,\alpha}$, we can then prepare the quantum feature state:
\begin{equation}\label{eq_feature_map}
    |\psi_{x, \alpha}\rangle = \hat{U}_{x,\alpha} |\psi_0\rangle = \Bigg( \prod\limits_{j=\text{odd}}^{N-1} \hat{U}_j(x_j) \prod\limits_{j=\text{even}}^{N-1} \hat{U}_j(x_j)\Bigg)^M \bigotimes\limits_{j=1}^N |\psi_j\rangle .
\end{equation}
The quantum circuit corresponding to this final feature map, as used in our experiments, is illustrated in Figure~\ref{fig_heis_ex} for 4 qubits and 3 features.

Finally, as the set of operator measurements $\{\hat{O}_i\}_{i=1}^q$, we choose $q$ number of $b$-local Pauli strings $\{\hat{S}_i^b\}_{i=1}^q$, where $b \in \{1, 2\}$. We always include all the $b$=1 strings, which correspond to single-qubit Paulis $\hat{P}_j \in \{\hat{X}_j, \hat{Y}_j , \hat{Z}_j\}$ on a qubit $j$ within our operator set. This means that on $N$ qubits we have at least $q = 3N$, while in some cases a few operators with $b=2$ are additionally included so that $q > 3N$. With this setup, element $i$ of the transformed feature vector $x'$ is given by 
\begin{equation}\label{eq_pqfm_features}
x'_i = \langle \psi_{x, \alpha} | \hat{O}_i | \psi_{x, \alpha} \rangle,
\end{equation}
where $\hat{O}_i$ is the $i^{th}$ Pauli string included in the measurement operator set.

Since the described procedure, as summarized in Equation~\ref{eq_pqfm}, is only a transformation of information contained in a single market event vector $x$ -- meaning $\phi$ adds no additional information, and is independent of market time and associated market event outcome label $y$ -- then it directly follows that also the same time- and label-independence translates to the final quantum-generated transforms $x'$, as obtained from Equation~\ref{eq_pqfm_features}. When repeating this procedure for all vectors $x$ in the originating financial time series $\mathcal{X}$, then also the resulting collection of vectors $x'$ becomes a time series $\mathcal{X}'$ that shares the originating stochastic nature of $\mathcal{X}$. This yields that the final dataset $\mathcal{D}'$, combining $\mathcal{X}'$ with the identical $y$ from $\mathcal{D}$, provides for trade execution learners no theoretical guarantees on prediction errors $\tilde{\varepsilon}_q$, as similarly elaborated for $\tilde{\varepsilon}_0$ in Section~\ref{subsec_tradelearning}. Therefore, any observed differences $| \tilde{\varepsilon}_q - \tilde{\varepsilon}_0 |$ in experimental tests discussed in the remainder of the paper are data-instance specific and cannot be explained by the theory introduced here.

\section{Empirical analysis setup}
\label{sec_setup}
The experiment is designed to test the methodology introduced in the prior section through an empirical analysis of actual algorithmic trading responses to corporate bond RFQs under real market conditions at each respective time instance. It uses a sample of real trading data by HSBC at production scale, the latest generation of an IBM quantum computer, and a cluster of classical computers for modeling. The details of the data composition, transformation, trade event sampling, and quantum feature generation on hardware are described in Section~\ref{subsec_data} and \ref{subsec_qfeaturegen}, and the adopted procedure for statistical learning, backtesting the trade execution likelihood estimates, and benchmarking various models in Section~\ref{subsec_backtesting}.

\subsection{Trading data}
\label{subsec_data}
The given data source sample corresponds to a trading period that ranges from September~1, 2023 to October~29, 2024 with 294~trading days of intraday algorithmic responses to overall 1~073~926 RFQs linked to 5166~bonds and 747~associated tickers, covering primarily the European corporate bond market. The time resolution of this tabular timeseries data is at the order of microseconds, where each row corresponds to a unique vector of features that encode a fixed choice of a market state representation that was observed at the time of an incoming RFQ, as defined in Section~\ref{subsec_staterep}. Accordingly, each feature vector, also referred to as the \textit{market state} or the \textit{trade event} vector at a unique time instance, is associated with a binary label that captures with the value "1" the successful acceptance of a submitted RFQ response or otherwise takes the value "0".

\begin{figure}[b!]
\centering
\includegraphics[width=\columnwidth]{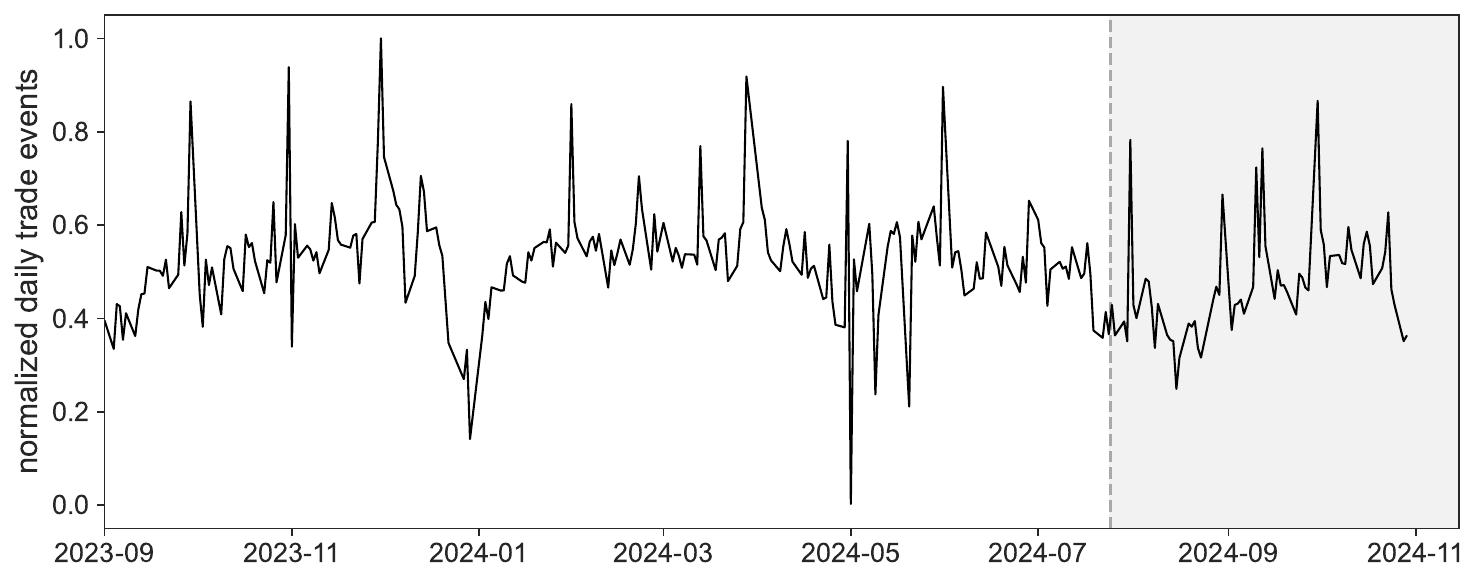}
\caption{Overview of the bond trading timeseries dataset used in this work, showing the number of daily RFQ events, normalized by its peak, over the market period from September 1, 2023 to October 29, 2024. The dashed line on July 24, 2024 and the gray-shaded region indicate the selected active window for the fill probability estimation analysis with respective events that incorporate available market and bond information from before.}
\label{fig_full_dataset_norm_trade_events_per_day}
\end{figure}
\begin{figure}[b!]
\centering
\includegraphics[width=\columnwidth]{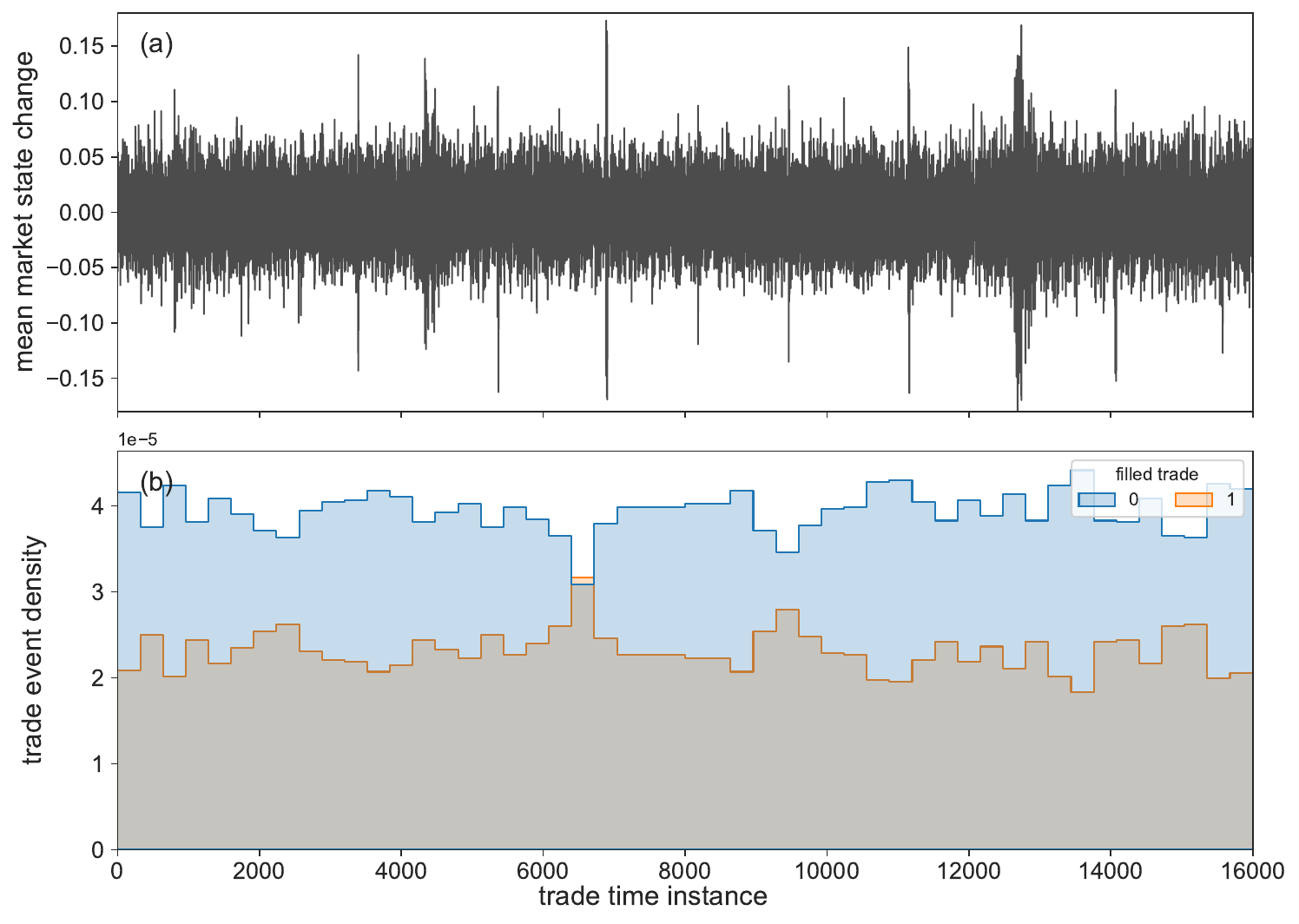}
\caption{Representative sample of 16~000 trade events with daily RFQ coverage, showing (a) market state and respective mean event vector changes between neighbouring time instances, and (b) normalized histogram of labels associated with these events, indicating executed or filled trades with “1”, or otherwise with “0”.}
\label{fig_batch12_classical_input_data_meanstatechange_targetstats}
\end{figure}
The event is constructed in continuous real market time with access to past information only, meaning technically up to a microsecond ago. It comprises an extensive set of multivariate timeseries data analyses capturing statistical characteristics linked to each specific bond at micro- to macro scales in both time and granularity dimensions: from RFQ, ticker to associated industry sector levels, and short- to long time lags that probe market dynamics and buy/sell-side responses. These characteristics are then represented in a dense set of 216~numerical features that form the event. We define in this experiment a confined trading window with events of interest and their respective rolling window of past information starting from July~24, 2024. This is shown in Figure~\ref{fig_full_dataset_norm_trade_events_per_day} with a dashed line and a shaded region representing the \textit{active trading time window} with 69~trading days used in this work. It also shows a varying daily algorithmic trading activity with irregular peaks.

In this active time window, we select only bonds that do not mature during that time, which amounts to 143~912~RFQs, linked to 3425 bonds and 652 tickers. Their respective events, as described above, also encode $\sim 9$ million intraday signals, such as price information. Each individual event feature value or groups of features, such as those differing only by time lags, are scaled between $-1$ and $+1$. We then randomly generate a representative sample of 16~000 events corresponding to RFQs on each day of this trading window. The sample is illustrated in Figure~\ref{fig_batch12_classical_input_data_meanstatechange_targetstats}, which exhibits its stochastic nature with the event vector changing from one to the next time instance in average by $\sim 5$\% with some temporal shocks, while the fraction of accepted RFQ responses with $\sim 37$\% remains relatively constant.

\subsection{Quantum feature generation and event matching}
\label{subsec_qfeaturegen}
Given a trade sample from the process above, the classical features are first normalized with the scaler \mbox{$\tilde{x}_i = (x_i - \mu)/w$,} where $\mu$ and $w$ are the mean and standard deviation of the sample, respectively. The normalized features are then further transformed with $2\pi*\tanh{\tilde{x}_i/3}$ to reduce the effects of outliers. This also ensures that these classical input feature values are in a suitable range for the subsequent quantum encoding as rotations, enabling controlling the full range of quantum gate rotation with a scaling parameter $\alpha$.

We then generate one projected quantum feature vector for each classical event vector, as described in Section~\ref{subsec_quantumfeature}, using a quantum hardware device. Since today's devices are still inherently noisy and may alter the generated feature outputs, we also use a noiseless quantum simulator running on a classical computer to compare with. The detailed procedure is explained in the following.

\subsubsection{Classical simulation of quantum}
\label{subsubsec_qsim}
The Matrix Product State simulation method~\cite{vidal2003efficient} and the backend of Qiskit Aer~\cite{qiskit2024} with its so-called Estimator Primitive under the default settings are used. The circuit consists of 109 qubits with the specific circuit parameters described in Table~\ref{tab_circ_params_hw_utility}, which also introduces the circuit labels "shorter" and "longer" as referred to later. The PQFM transform of the classical event vector is then extracted by taking the set of $1$-local Pauli observables for measurement as described in Section~\ref{subsec_quantumfeature}. Since this corresponds to the $X$, $Y$, and $Z$ observable measurements on each qubit, the total number of PQFM output features, $q$, is $3N = 3 \times 109 = 327$.

\subsubsection{Quantum hardware execution}
\label{subsubsec_qhw}
\begin{figure}[t!]
\centering
    \includegraphics[width=\columnwidth]{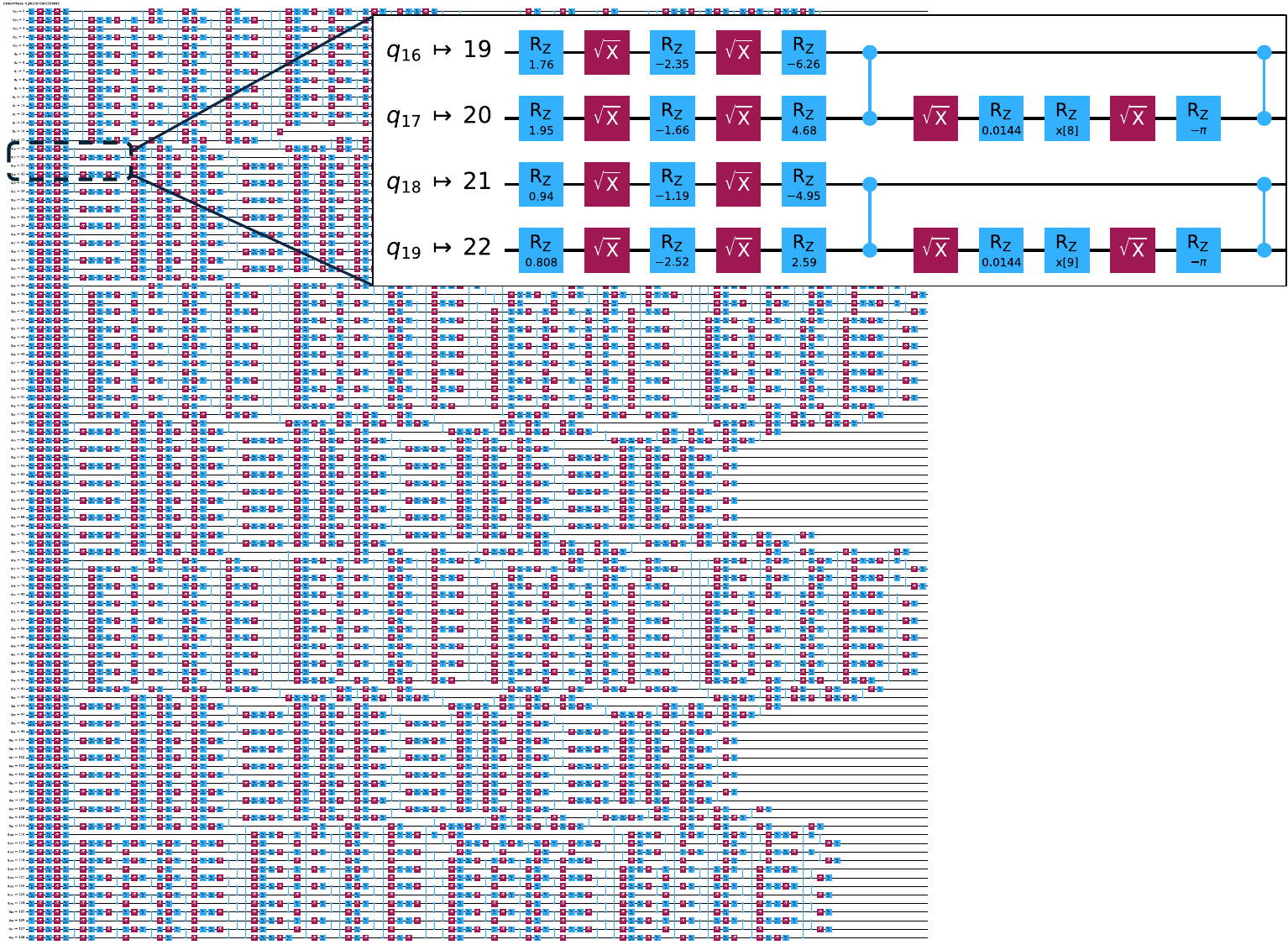}
    \caption{
        Example of the quantum circuit for the "shorter" Heisenberg feature maps with 109 qubits produced by the Qiskit transpiler targeting the \texttt{ibm\_torino} quantum hardware device.
    }
\label{fig_compiled_circuit_shorter}
\end{figure}

The same approach as for simulation and parameter settings in Table~\ref{tab_circ_params_hw_utility} is used for the quantum hardware experiments. These are primarily conducted on the IBM Quantum system \texttt{ibm\_torino}, which leverages a Heron r1 processor with 133 fixed-frequency transmon qubits that are connected via tunable couplers on a heavy-hex lattice~\cite{abughanem2025ibm, IBMQuantum_processorTypes}. The expectation value of each observable is computed using 4096 shots. Since the quantum hardware is susceptible to noise, many methods to mitigate and/or suppress the impact of hardware noise are available. In this study, we apply two error mitigation methods: Pauli Twirling~\cite{wallman2016noise, geller2013efficient} for error suppression, and Twirled Readout Error eXtinction (TREX)~\cite{chen2021robust, van2022model} for measurement error mitigation through the Qiskit IBM Runtime Estimator primitive~\cite{IBMQuantum_qiskitPrimitives}. An example of the compiled quantum circuit for \texttt{ibm\_torino} is shown in Figure~\ref{fig_compiled_circuit_shorter}.
\begin{table}[t!]
    \centering
    \begin{tabular}{llr} \hline
        Quantum circuit name & Parameter & Setting \\ \hline
        Heisenberg "shorter" & qubits & 109 \\
        & blocks & 1 \\
        & alpha & 1.0 \\
        & initial state & random \\
        & random seed & 1 \\ \hline
        Heisenberg "longer" & qubits & 109 \\
        & blocks & 2 \\
        & alpha & 0.1 \\
        & initial state & random \\
        & random seed & 0 \\ \hline
    \end{tabular}
    \caption{Quantum circuit parameter settings used for both noiseless quantum simulator and hardware experiments. The term "blocks" refers to the number of repetitions of encoding all classical input features in the event vector. Since there are 216 features, one block corresponds to two repetitions or trotter steps $M$ as described in Section~\ref{subsec_quantumfeature}. The parameter $\alpha$ is the shared rotational scaling, and the random seed is set to generate the random initial state for each qubit sequentially.}
    \label{tab_circ_params_hw_utility}
\end{table}

\subsubsection{Classical-quantum event matching} 
\label{subsubsec_cqem}
Based on the methodology introduced in Section~\ref{subsec_staterep} and \ref{subsec_quantumfeature}, we further leverage the fact that the above quantum feature generation procedure is only a functional transformation, not an informational extension, of classical data features that encode a normalized market event with unknown outcome, given its underlying stochastic process it is sampled from. This means that if an event at time $t_i$ is observed again at another arbitrary time $t_j \neq t_i$, the previously observed outcome label $y_i$ does not determine $y_j$, which instead may equally likely take the opposite outcome.

\begin{figure}[t!]
\centering
    \includegraphics[width=\columnwidth]{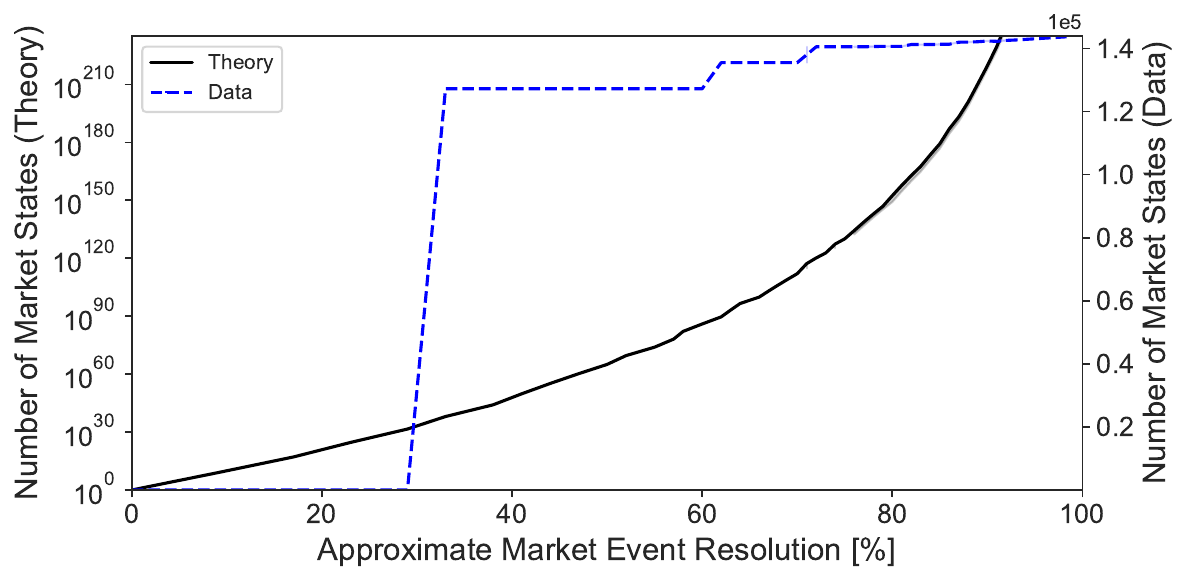}
    \caption{
        Number of theoretically representable classical market states for 216 normalized features as a function of event matching resolution, compared to the number of realized states in our trading dataset.
    }
\label{fig_number_market_states_theory_vs_data_sample}
\end{figure}

Motivated by this outcome or label inference independence of recurring events and the direct link of classical to quantum-transformed event vectors, we introduce a simple but robust event matching technique to reuse quantum-generated features for unseen but similar classical events that were used to generate them with a quantum circuit. The idea is to generate an event-level identifier $\kappa$ through discretization of normalized classical events $x$ from a large population $X$, from which we then draw a sample $X_{\text{sample}} \subset X$ to produce quantum-generated features, and then reuse these features for other classical event instances $x_j \in  X \setminus X_{\text{sample}}$ that share the same $\kappa$ of $x_j \in  X_{\text{sample}}$. While any kind of discretization function could be used, a numerical binning over the fixed range $[-1, +1]$ of event values in our trading dataset is sufficient. While a non-linear binning scheme, optimized for the particular event distribution, could be derived, given the purpose of our experiments, we choose a basic linear scheme with uniformly distributed bins of equal size and adopt the following classical-to-quantum event matching protocol:
\begin{enumerate}
    \item Take the pre-processed and scaled dataset of the full active trading window $X$.
    \item Split the classical event feature value range into $N_{\text{bins}}$ number of bins, record bin indices in which the respective feature values fall, and orderly concatenate into a $\kappa$ identifier per event vector to derive a $\kappa$-labeled set of event tuples, $X^\kappa := \{ (\kappa, x)~|~\forall x \in X \}$.
    \item Generate PQFM vectors using a quantum circuit for a subset of classical event vectors $X^\kappa_{\text{sample}} \subset X^\kappa$.
    \item Match $\kappa$'s from $X^\kappa_{\text{sample}}$ with those from $X^\kappa \setminus X_{\text{sample}}$ to compile $X^\kappa_{\text{match}}$ and associate PQFM vectors, respectively.
    \item Apply a $\kappa$-grouped PQFM unification function $\Gamma^\kappa$ to those PQFM vectors linked to $X^\kappa_{\text{match}}$ to assure unique classical to quantum-generated feature vector associations.
\end{enumerate}
In our experiments, we use the mean function for $\Gamma^\kappa$, keep the order of features in an event vector always fixed, and scan with a defined set of $N_{\text{bins}} \in \{ 4, 10, 30, 60 \}$ that corresponds to a range of bin sizes covering $[0.0\bar{3}, 0.5]$. This can also be thought of as the sensitivity of changes in classical feature values. But for convenience, we refer to $(1 - 1/N_{\text{bins}})$ as the event matching resolution. Figure~\ref{fig_number_market_states_theory_vs_data_sample} shows how this approach would allow representing an exponentially large number of market states with $(N_{\text{bins}})^{216}$ in theory, while our trading dataset is obviously \mbox{capturing} only real and realized states in a confined market environment, which, with increasing $N_{\text{bins}}$, slowly converges towards the number of samples in $X^\kappa$, as expected.

\subsection{Trade execution backtesting and benchmarking}
\label{subsec_backtesting}
Any potential performance gain from hypothetically embedding a fill probability estimator that has access to classical and quantum-generated features, as illustrated in Figure~\ref{fig_bigpicture}, can only be assessed indirectly, since trading behavior changes as a consequence of different fill probability estimates may lead to unknown responses from the market that influence the conditions for the next time instance and respective RFQs. Nevertheless, the method of backtesting, which is widely used in model validation in the financial industry, lets us relatively compare and benchmark different models while keeping the test environment fixed. In this paper, we are only concerned about any observable performance differences when changing the input data with fixed model parameters and not about the accuracy of their magnitude, which would require more sophisticated calibrations and statistical hypothesis tests with associated uncertainties. We leverage the following backtesting protocol for the compiled dataset described in Section~\ref{subsec_data}:
\begin{enumerate}
    \item At time instance $t_k$ and respective arrival of RFQ $k$, compose a training dataset with event vectors of type $\eta$ for the past trading time period $\mathcal{T} := [t_{k - b}, t_{k - a}]$ with $a > 0$ and $b > a$.

    \item Given a binary classification model of type $m$ with associated hyperparameters, train model variations for different parameter settings and cross-validate on random subsets of events in $\mathcal{T}$ with associated RFQs and labels $y \in \{0, 1\}$ that mark respective RFQ outcomes.

    \item Take the best learned model that maximized a chosen performance metric for randomized in-sample testing within $\mathcal{T}$ and test on the out-of-sample event vector associated with RFQ $k$ at $t_k$ by estimating its fill probability, where the time difference $\Delta_{ka} := t_k - t_{k - a}$ is referred to as the \textit{blinding window} to prior market information.

    \item Iterate index $k$ to next evaluation instance and repeat from step 1 above.
\end{enumerate}

This protocol is executed over all permutations of defined values that $\eta$, $\mathcal{T}$, $\Delta_{ka}$, and $m$ can take, as specified in \ref{app_1}. Practically, step 3 is used to estimate the fill probability not only for instance $t_k$ but for all following instances that fall into a time window of interest, such as one or two trading days into the future. This can then be used to measure the sensitivity of the estimator to blinding prior market information, and it informs how frequently a model would need to be retrained before its performance degrades below a given threshold. In the experiment reported in this paper, we consider $\eta$ as the benchmarking flag that feeds a given model of type $m$ with either classical or quantum-generated data, while the rest remains identical. Based on evaluation of an extensive set of basic to more modern and advanced statistical learning algorithms, our final choice of model types $m$ in this paper aligns with those more commonly encountered in the industry:
\begin{itemize}
    \item Logistic Regression (LR),
    \item Gradient Boosting (XGB),
    \item Random Forest(RF), and
    \item Feed-Forward Neural Network (NN).
\end{itemize}
The first corresponds to a basic linear model, while the others belong to the class of non-linear models. This choice is further motivated by keeping the model complexity as low as possible while adhering to the experiment's purpose noted above.

As the primary model performance metric used for model training, testing, and relative comparisons between models, we select the area under the receiver operating characteristic curve, thereafter referred to as AUC metric. It corresponds to a measure of probability that the trained model ranks a randomly chosen positive $y = 1$ RFQ response instance higher than a randomly chosen negative $y = 0$ instance. This metric is well-suited for our purpose since, for slightly to moderately imbalanced data, as is the case in our trading dataset with $\sim 37$\% positive labels, it is often considered the gold standard metric and commonly used and preferred over accuracy. This is because accuracy can be misleading with imbalanced data, biased towards predicting the majority class, and failing to capture the true discriminative power of a model. AUC, on the other hand, since it evaluates the relative ranking of positive vs. negative instances, is robust to skewed class ratios and also independent of the particular threshold used for assigning labels, thus decoupling the modeling from probability calibration~\cite{bradley1997use, fawcett2006introduction}.

\section{Results}
\label{sec_results}
We report our observations from various statistical learning experiments that aim to estimate the fill probability of corporate bond RFQs using the methodology defined in Section~\ref{sec_method}, and its implementation using the backtesting setup for a given algorithmic trading data sample defined in Section~\ref{sec_setup}.

\subsection{Quantum feature distributions}
\label{subsec_qfeaturedists}
Here we first analyze and compare the distributions of the classical and quantum-generated features, before presenting the results that used them in backtesting experiments in the remainder of this section.

\begin{figure}[t!]
  \centering
  %
  \begin{subfigure}[b]{\columnwidth}
    \centering
    \includegraphics[width=0.24\columnwidth]{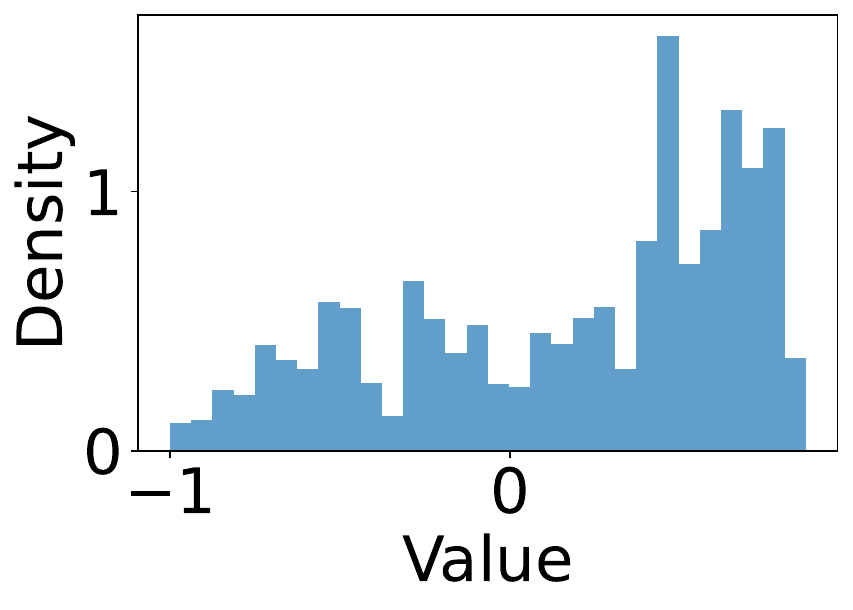}
    \includegraphics[width=0.23\columnwidth]{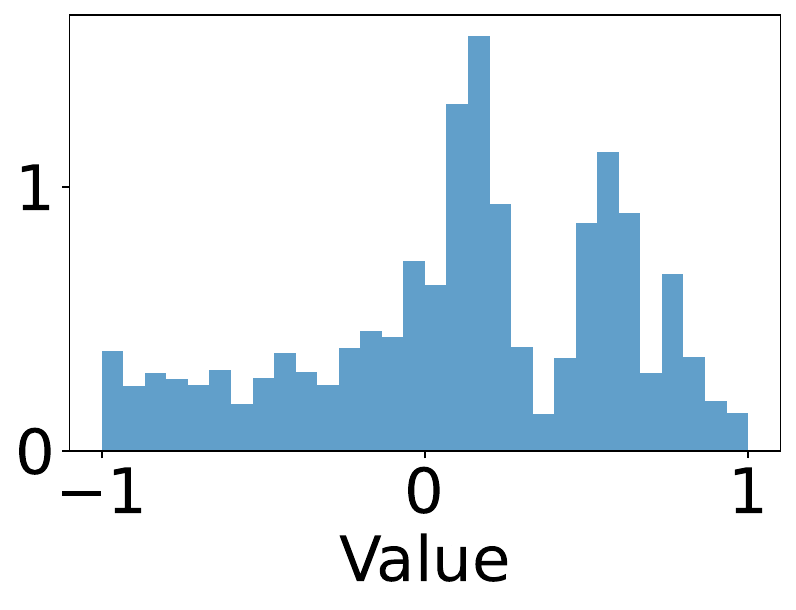}
    \includegraphics[width=0.25\columnwidth]{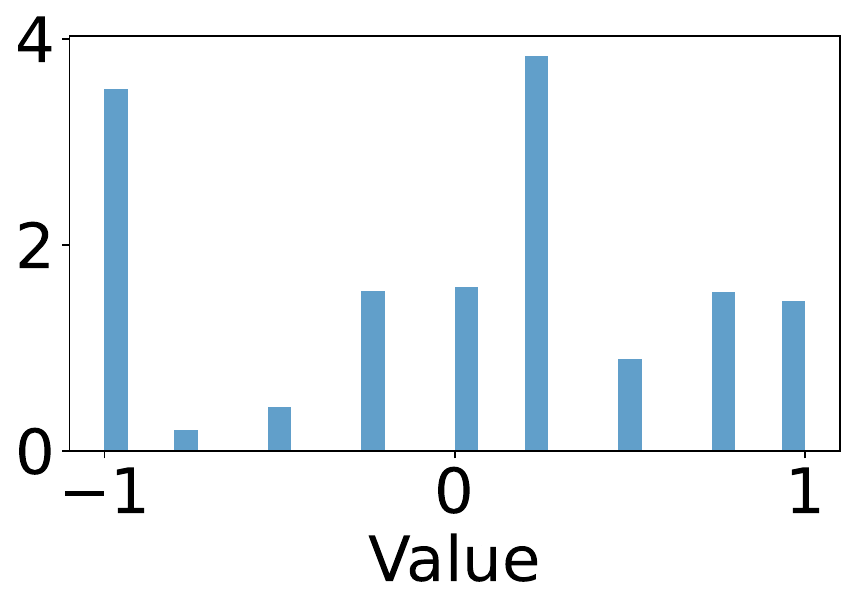}
    \includegraphics[width=0.24\columnwidth]{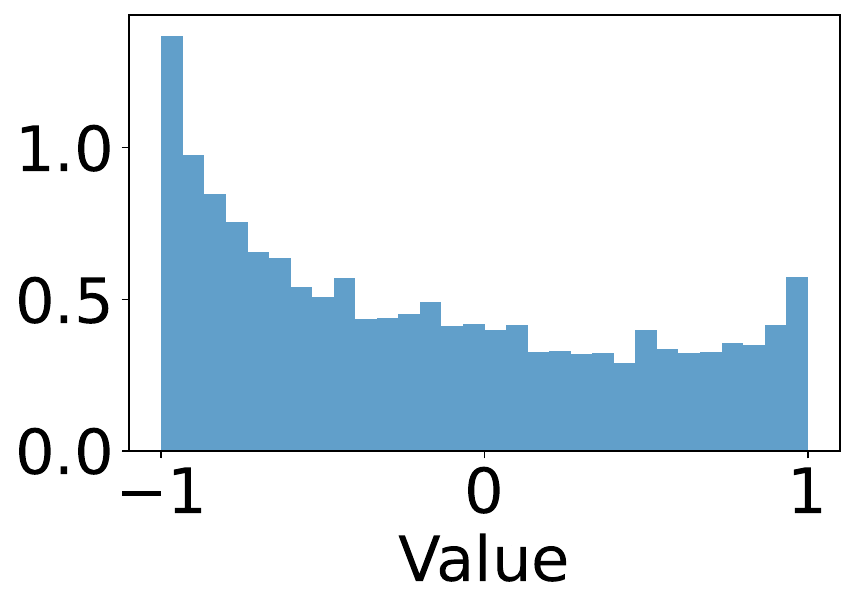}
    \caption{Classical feature histograms: from left to right features 98, 177, 2, 19}
  \end{subfigure}
  \vspace{0.2em}
  
  \begin{subfigure}[b]{\columnwidth}
    \centering

    \includegraphics[width=0.24\columnwidth]{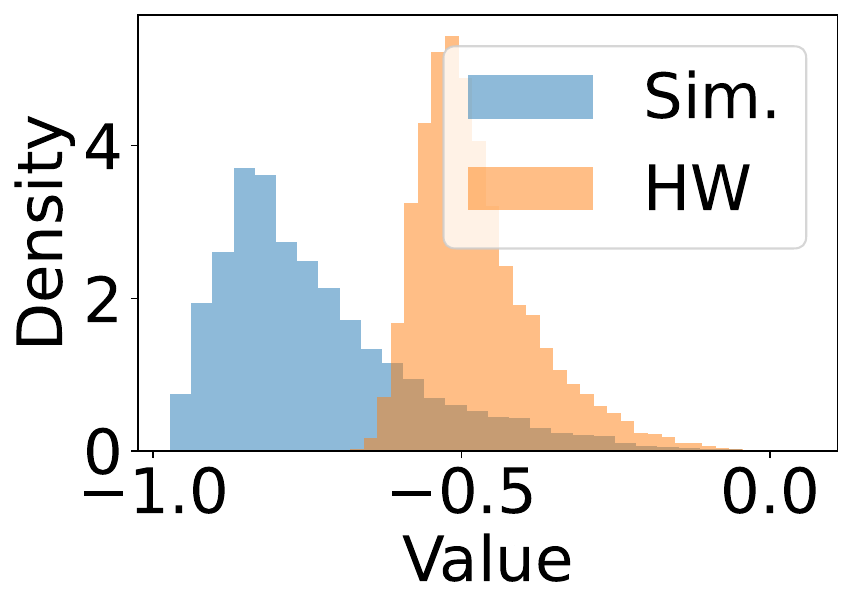}
    \includegraphics[width=0.25\columnwidth]{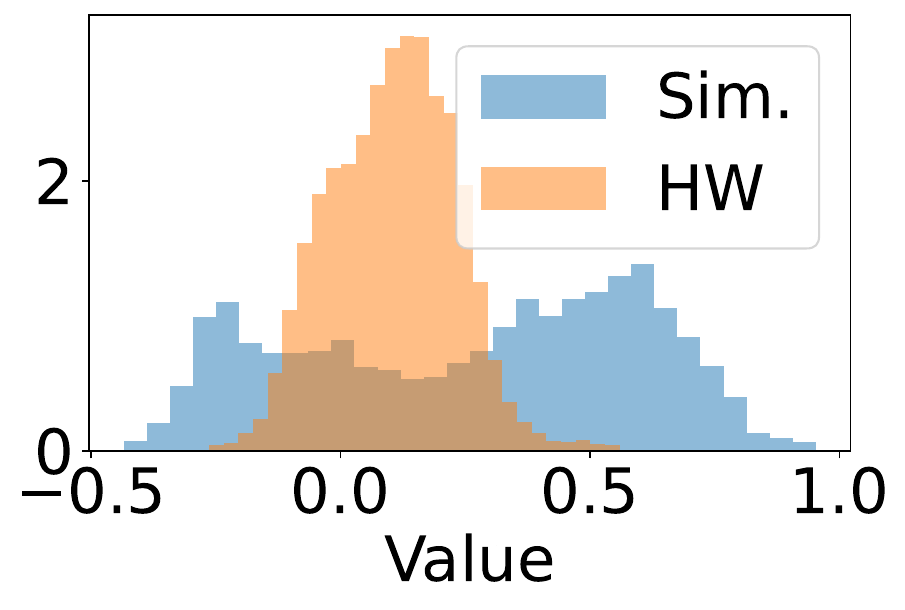}
    \includegraphics[width=0.23\columnwidth]{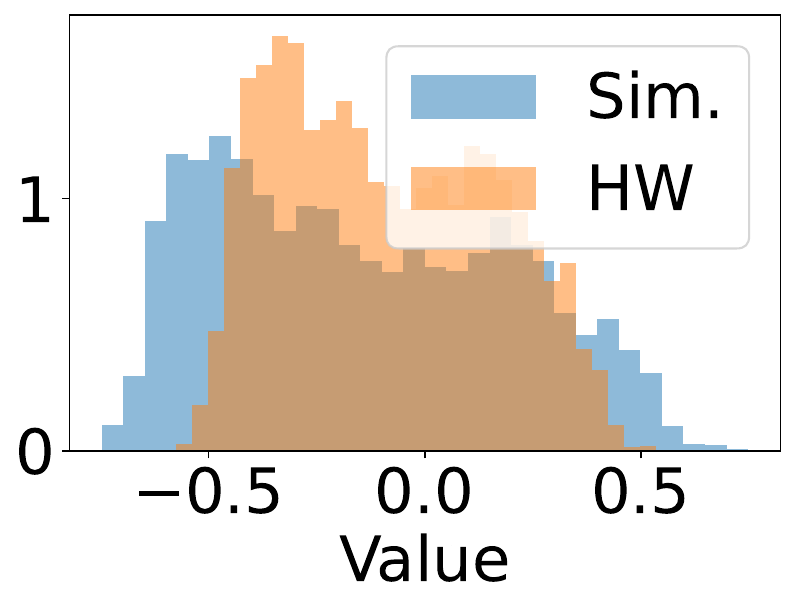}
    \includegraphics[width=0.23\columnwidth]{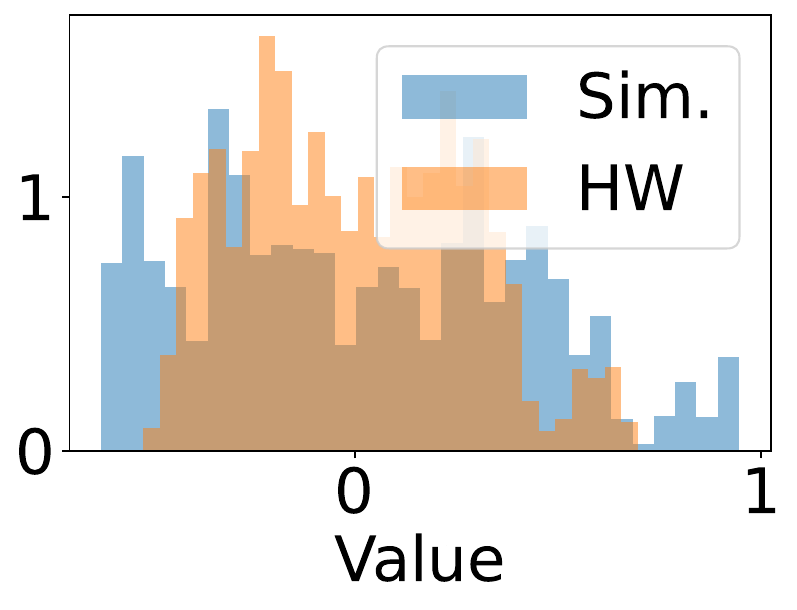}
    \caption{Shorter circuit PQFM observable histograms: YQ90, XQ18, YQ7, ZQ80}
  \end{subfigure}
  \vspace{0.2em}
  
  \begin{subfigure}[b]{\columnwidth}
    \centering
    \includegraphics[width=0.24\columnwidth]{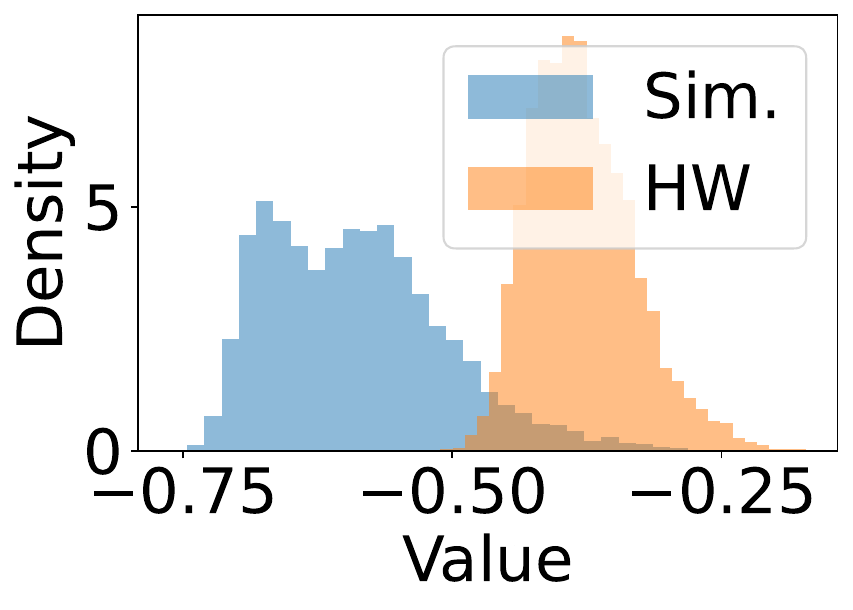}
    \includegraphics[width=0.23\columnwidth]{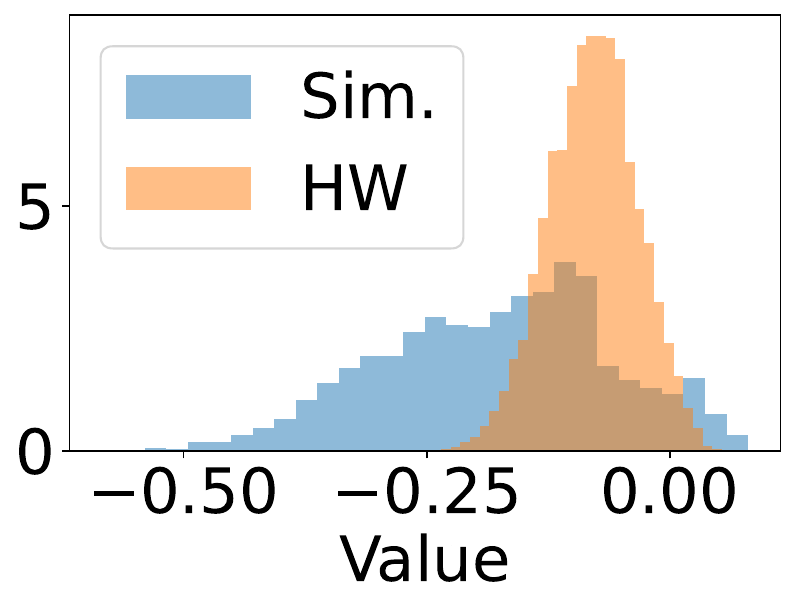}
    \includegraphics[width=0.23\columnwidth]{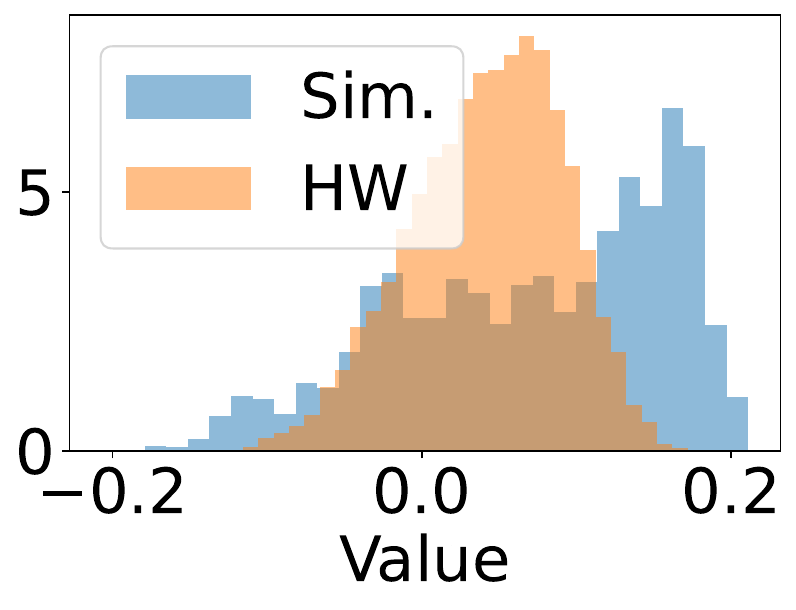}
    \includegraphics[width=0.24\columnwidth]{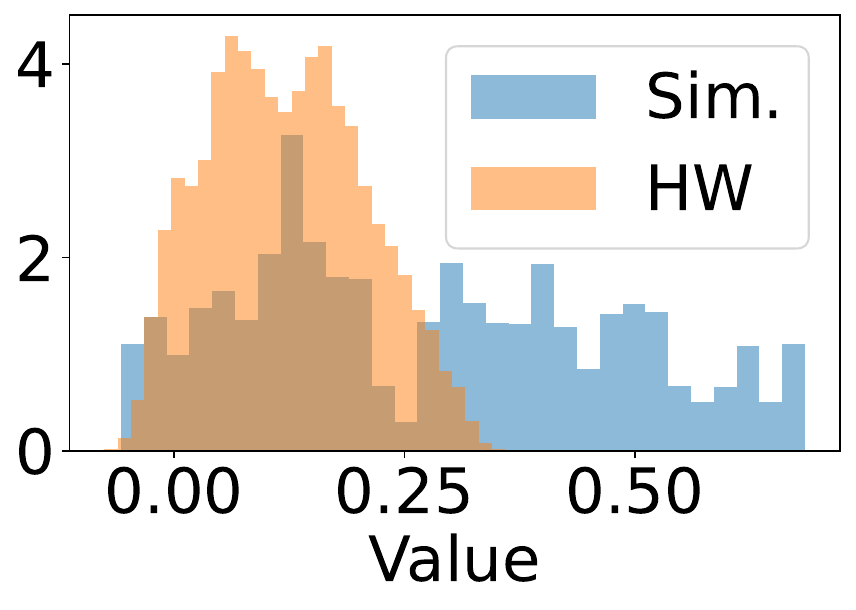}
    \caption{Longer circuit PQFM observable histograms: YQ90, XQ18, YQ7, ZQ80}
  \end{subfigure}
  \caption{Representative example feature distributions.  Histograms of select representative classical feature values (first row) and quantum-generated features, for the shorter circuit (second row) and longer circuit (third row) -- for both noiseless simulation (Sim.) and hardware runs (HW). The first 2 columns are representative of the most common distributions, with the last 2 showing more variety. Quantum features are the same for the last 2 rows, while classical features are not directly related (arbitrarily selected). Quantum features are referred to by the observable and qubit: for example, ``YQ90'' means the Y observable of qubit 90.}
  \label{fig_feature_distributions}
\end{figure}
Histograms for a select set of representative features are shown in Figure~\ref{fig_feature_distributions}, for both classical (first row) and quantum-generated features (bottom two rows). The examples of the first two columns capture the most common feature distributions, while some additional variety is illustrated with the last two columns. The quantum-generated feature distributions are shown for both "shorter" and "longer" Heisenberg circuits with the same quantum feature in each column, and for both noiseless simulation using a classical computer and hardware using the \texttt{ibm\_torino} device.

We observe the classical features to generally have complex, multi-modal, and non-smooth distributions with some exceptions. Among these are discrete or partially smooth but mostly skewed cases. However, the quantum-generated features are overall smoother and less complex in general (first two columns), with rather few cases of less-smooth distributions, in particular for noiseless simulation (last two columns).

This illustrates that the quantum transform is creating significantly different features from classical ones, with generally smoother and more normal distributions. More generally, as we move from noiseless simulation to noisy quantum hardware, we observe additional smoothing and normalization of the feature distributions, and similarly when shifting from the "shorter" to the "longer" circuit. This seems to indicate that the more quantum noise effects are present, such as additional circuit operations and decoherence time in the "longer" circuit, the smoother and more concentrated towards 0-mean are the resulting quantum feature distributions -- as reflected in the hardware results.

\subsection{Trade backtesting analysis}
\label{subsec_resbacktesting}
Based on the four selected machine learning models with either classical or quantum-generated features as data inputs and the trade execution backtesting protocol from Section~\ref{subsec_backtesting} for the active trading window defined in Section~\ref{subsec_data}, we present our empirical observations in this section.

\begin{figure}[t!]
\centering
\includegraphics[width=\columnwidth]{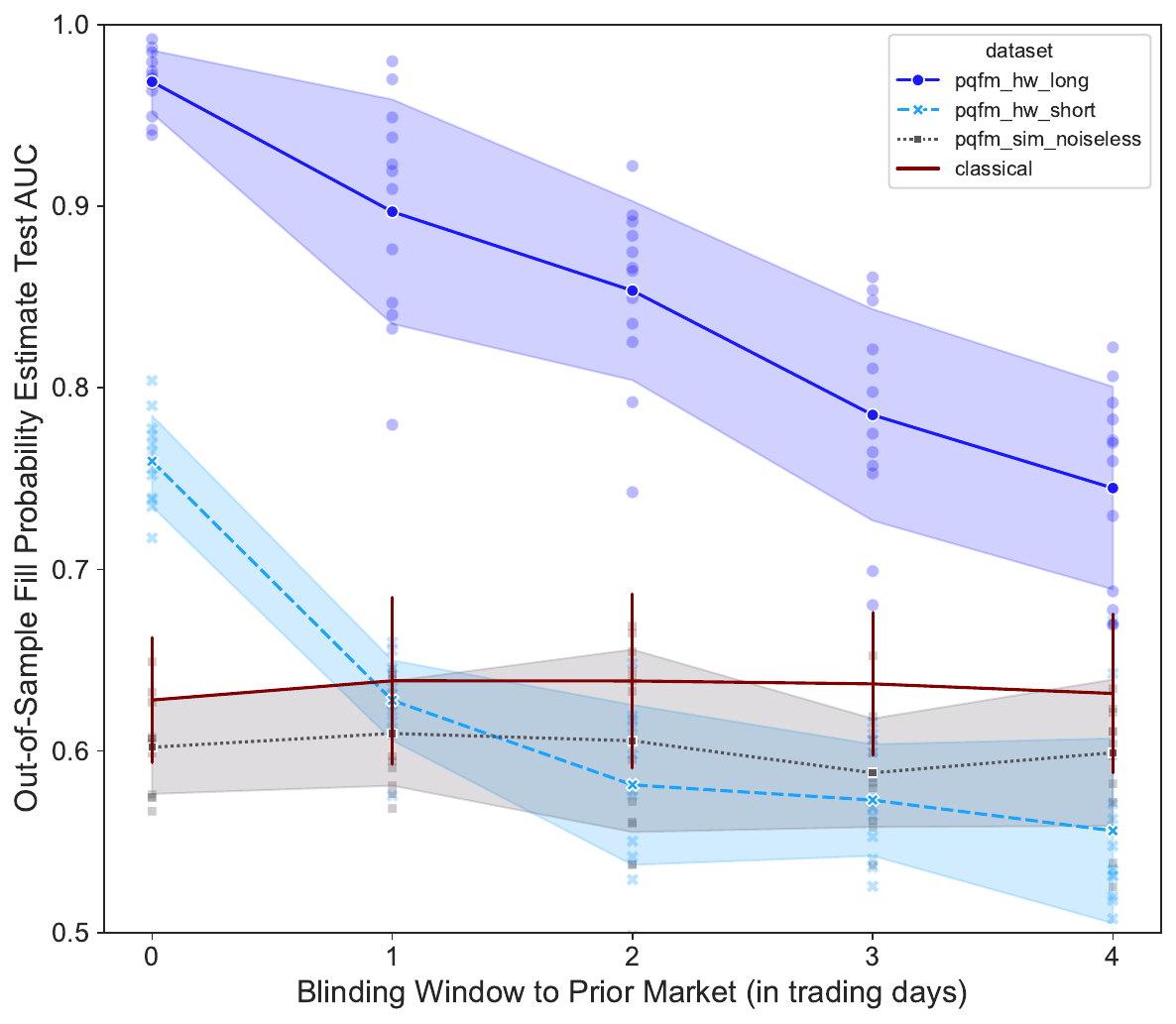}
\caption{Test AUC scores and associated standard deviation bands for out-of-sample fill probability estimates from trade backtesting LR, XGB, RF, and NN models with different data inputs as a function of blinding prior market information. In this case, "0" blinding refers to deltas of $\geq 1\mu s$ to $< 24$h between training data and trade test samples in the future, followed by "1" trading day that moves this window by 24h, and so on.}
\label{fig_batch12_multimodel_timeseries_probeval_c_qsim_qhw_qhwshort_benchmark_combined}
\end{figure}

Figure~\ref{fig_batch12_multimodel_timeseries_probeval_c_qsim_qhw_qhwshort_benchmark_combined} summarizes the results of observed test AUCs for out-of-sample fill probability estimates as a function of blinding prior market information from the learning models. This shows a linear flat profile for models with classical inputs and an average test AUC of $\sim 0.63 \pm 0.04$ over up to 4 trading days of blinding, which may be interpreted as a lack of time dependence with no sensitivity to temporal trading signals in the adjacent market environment before the arrival of RFQs. But, on the contrary, there is also no evidence of whether and how frequent such signals occur, and if they can be learned from our chosen market state representation.

\begin{table*}[t]
    \centering
    \begin{tabular}{lllll} \hline
        Dataset source & Model & Median Test AUC & Median Test AUC  & Diff. to classical \\
                       &       & 0d              & 1d               & 0d / 1d [\%]          \\ \hline
        
        Noiseless quantum simulation
        & All
        & $0.60 \pm 0.03$
        & $0.60 \pm 0.03$
        & $- 3$ / $- 4$ \\
        
        & LR
        & $0.61 \pm 0.01$
        & $0.64 \pm 0.02$
        & $- 1$ / $+ 2$ \\

        & XGB
        & $0.59 \pm 0.03$
        & $0.61 \pm 0.03$
        & $- 6$ / $- 5$ \\

        & RF
        & $0.59 \pm 0.02$
        & $0.59 \pm 0.02$
        & $- 7$ / $- 7$ \\

        & NN
        & $0.58 \pm 0.02$
        & $0.59 \pm 0.01$
        & $\pm 0$ / $+ 1$ \\

        Quantum hardware
        & All
        & $0.75 \pm 0.02$
        & $0.62 \pm 0.03$
        & $+ 12$ / $- 2$ \\

        \small ("shorter" circuit)
        & LR
        & $0.75 \pm 0.02$
        & $0.61 \pm 0.02$
        & $+ 13$ / $- 1$ \\

        & XGB
        & $0.77 \pm 0.02$
        & $0.63 \pm 0.03$
        & $+ 12$ / $- 3$ \\

        & RF
        & $0.75 \pm 0.02$
        & $0.64 \pm 0.03$
        & $+ 9$ / $- 2$ \\

        & NN
        & $0.74 \pm 0.02$
        & $0.62 \pm 0.03$
        & $+ 16$ / $+ 4$ \\

        Quantum hardware
        & All
        & $0.97 \pm 0.02$
        & $0.88 \pm 0.06$
        & $+ 34$ / $+ 24$ \\

        \small ("longer" circuit)
        & LR
        & $0.95 \pm 0.03$
        & $0.81 \pm 0.06$
        & $+ 33$ / $+ 19$ \\

        & XGB
        & $0.96 \pm 0.02$
        & $0.89 \pm 0.06$
        & $+ 31$ / $+ 23$ \\

        & RF
        & $0.98 \pm 0.01$
        & $0.93 \pm 0.03$
        & $+ 32$ / $+ 27$ \\

        & NN
        & $0.97 \pm 0.02$
        & $0.86 \pm 0.04$
        & $+ 39$ / $+ 28$ \\
        
        \hline
        \end{tabular}
    \caption{Median Test AUC scores for out-of-sample fill probability estimates from trade backtesting LR, XGB, RF, and NN models using PQFM features from noiseless simulation or quantum hardware, showing results for no blinding (0d) and 1 trading day (1d) of prior market information and respective differences to the same models using classical feature inputs.}
    \label{tab_batch12_multimodel_timeseries_probeval_benchmark}
\end{table*}
A similar linear flat profile is found for models with quantum-generated inputs from noiseless simulations with an average test AUC of $\sim 0.60 \pm 0.03$ over up to 4 trading days of blinding, and therefore slightly below those with classical data inputs. As we move to models with inputs generated by quantum hardware, we find elevated model performance decay profiles that respond to blinding time. From an algorithmic trading and economic perspective, these decays may make sense, but as noted above, they cannot be interpreted as confirming the existence of temporal signals or evidence for having found them in quantum-transformed data to improve fill probability estimates. The test AUC elevations seem to differ by the relative amount of noise the respective PQFM features have been exposed to in the process. The one using the "shorter" Heisenberg circuit shows a median test AUC of $\sim 0.75 \pm 0.02$ without blinding, before decaying with a curvature to similar levels of the noiseless simulation results. As the noise exposure is increased with the "longer" circuit, we observe a significant jump to a median test AUC of $\sim 0.97 \pm 0.02$ that decays linearly with similar standard deviations up to 4 trading days of blinding. The two performance curves of models with features from the "shorter" and "longer" circuits do not show a constant shift between them. The results at a more granular level are found in Table~\ref{tab_batch12_multimodel_timeseries_probeval_benchmark}, which also shows the differences between all four models. Overall, a model performance gain of up to $\sim 34$\% over all models is observed, indicating substantially lower fill prediction errors with these quantum-generated features: $\tilde{\varepsilon}_q < \tilde{\varepsilon}_0$.

Despite this being a data-specific empirical analysis that is subject to the heuristic method we are using, and thus without any generalization guarantees, the magnitude of the time-dependent out-of-sample AUC uplifts seems questionable and cannot be explained in detail. In particular, the exact role that the intrinsic noise during the hardware execution of quantum circuits plays in these finally derived model performance gains is not understood.

Additional reproducibility tests with the same data and setup on other IBM Quantum systems using a Heron r2 processor, the \texttt{ibm\_fez} and \texttt{ibm\_marrakesh} devices, show the same effects as on the \texttt{ibm\_torino} device. For example, the median test AUCs without blinding are identical with up to $\sim 1$\% difference. As preliminary steps to further probe these effects, select variations of the "longer" circuit after transpilation are tested with respect to fill prediction errors, including the replacement of pairwise qubit operations with idle operations that preserve the decoherence times, for which comparable results on quantum hardware are observed. This clearly suggests that the exact form of the feature map in Equation~\ref{eq_feature_map} cannot be used to claim any theoretical understanding of these experimental effects. Also, further attempts through quantum simulations with artificially induced noise do not reproduce the beneficial quantum hardware results, thus demanding further investigation.

\begin{figure}[t!]
\centering
\includegraphics[width=\columnwidth]{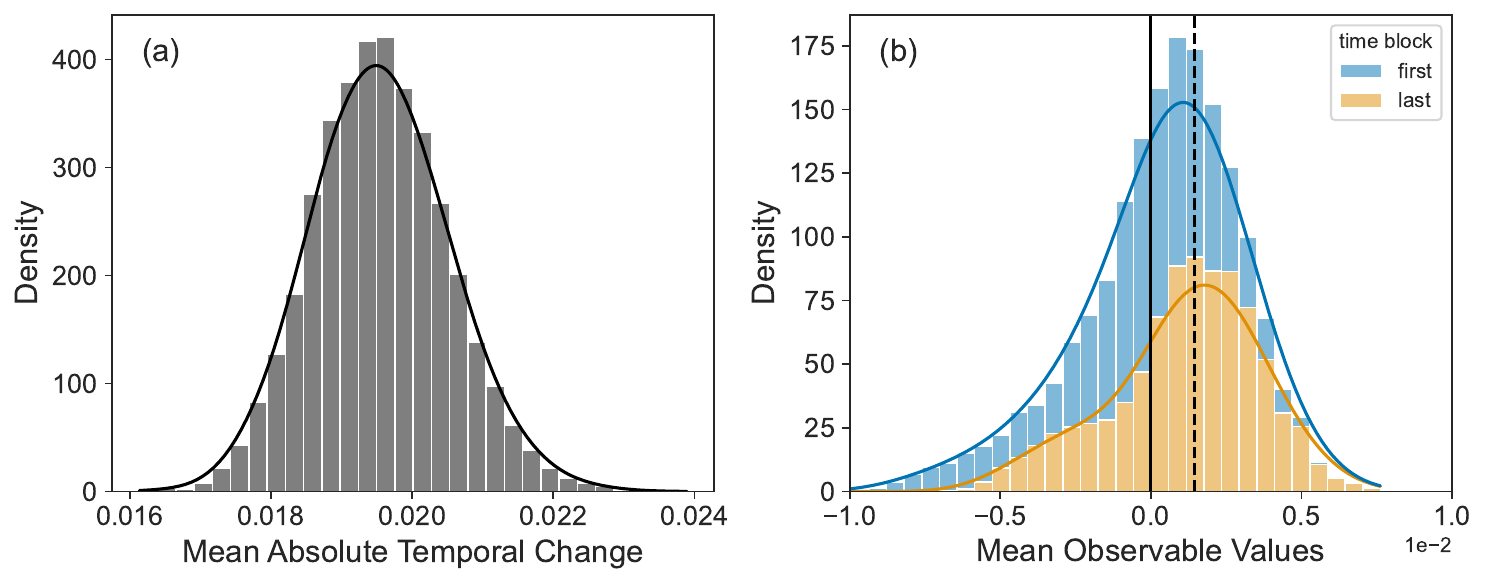}
\caption{Example of noise-induced drifts of PQFM features generated with the "longer" circuit on the \texttt{ibm\_torino} device with 16k identical input events: (a) shows the mean absolute change of observables from one to the next processed input event; (b) shows the distributions of mean observable values for the first and last third of processed events, with a drift of their medians by $\sim 1.4 \cdot 10^{-3}$, indicated by a dashed line.}
\label{fig_temporal_noise_drift_experiment}
\end{figure}
However, we cannot exclude that the order of processing these quantum circuits on hardware, and potential parallel temporal changes of the noise during processing, leave different characteristic time-dependent traces of noise in the resulting PQFM features. A simple experiment of running the "longer" circuit with identical classical event vectors as data input many times sequentially, as shown in Figure~\ref{fig_temporal_noise_drift_experiment}, indicates very small temporal changes to the resulting observables and respective drifts of the distributions during processing. Even if these hardware noise effects, despite drift corrections, would yield some form of non-trivial information leakage with respect to the RFQ outcome labels associated with event vectors, the usefulness of the generated PQFM features can still be probed by further decoupling the labels through the previously introduced classical-quantum event matching technique.

\subsection{Classical-quantum event matching}
\label{subsec_reseventmatching}
Motivated by the open questions of the out-of-sample trade backtesting results, we present here the results of reusing the quantum-generated features from the \texttt{ibm\_torino} device for unseen trade instances and thus their respective unseen labels.

\begin{figure}[t!]
\centering
\includegraphics[width=\columnwidth]{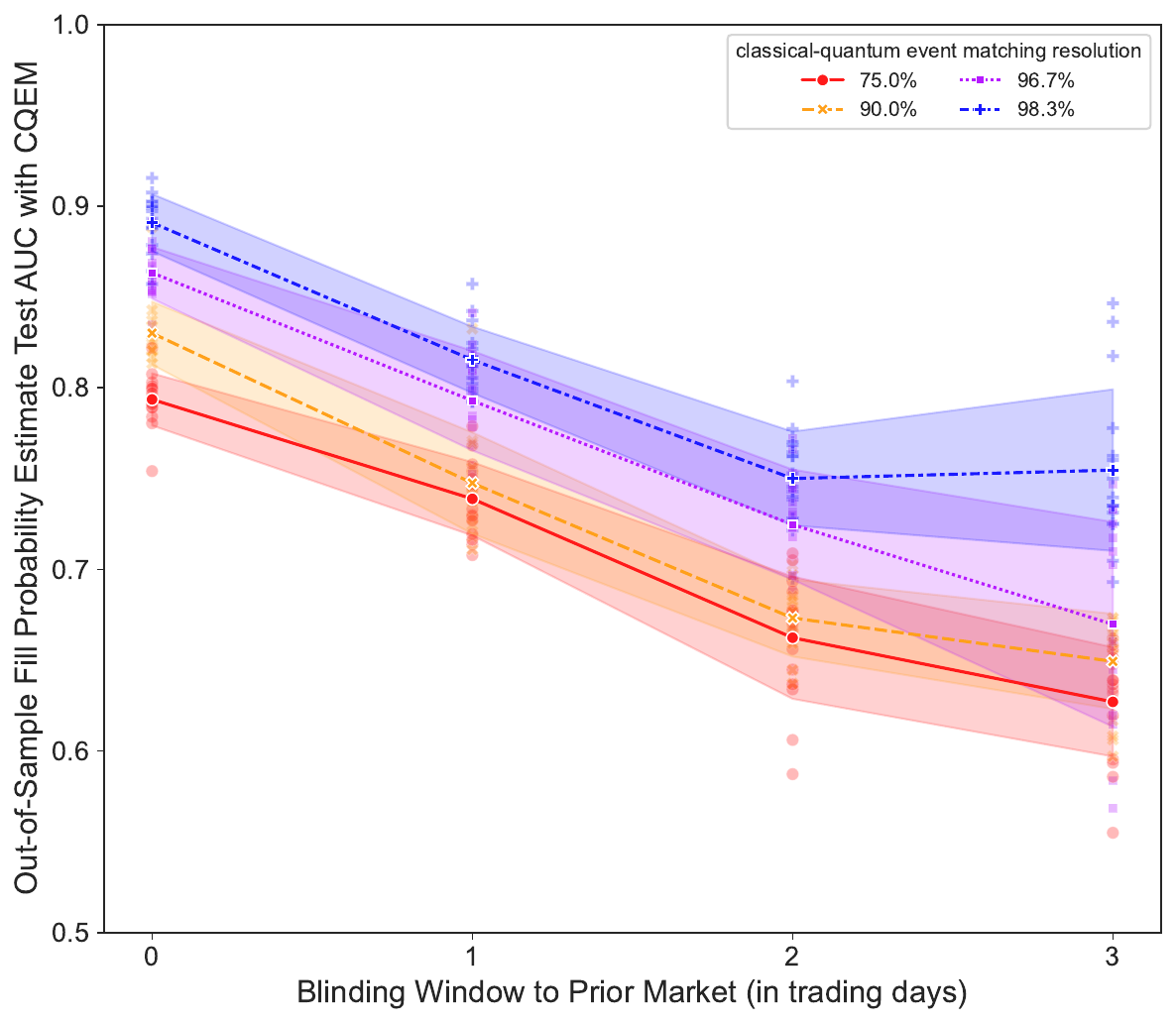}
\caption{Test AUC scores and associated standard deviation bands with classical-quantum event matching (CQEM) for out-of-sample fill probability estimates from trade backtesting LR, XGB, RF, and NN models with PQFM features generated by the "longer" circuit on the \texttt{ibm\_torino} device, as a function of blinding prior market information.}
\label{fig_batch12_multimodel_timeseries_probeval_classical_event_matched_qhw_benchmark_combined}
\end{figure}
We use the event matching protocol defined in Section~\ref{subsubsec_cqem} in combination with the trade execution backtesting protocol defined in Section~\ref{subsec_backtesting}. This leverages the same sample of 16k PQFM features from the "longer" circuit used in Section~\ref{subsec_resbacktesting}, but only for training the models. It then estimates fill probabilities for future trade instances from the full active trading window, but excluding these 16k training events, which leaves 127~912 events for matching.

Although statistically limited by the relatively small number of generated PQFM features with respect to the full trading dataset, the observed out-of-sample test AUCs in Figure~\ref{fig_batch12_multimodel_timeseries_probeval_classical_event_matched_qhw_benchmark_combined} for different event matching resolutions still show the same characteristic linear decay profiles as before. The chosen resolution controls the model performance, with the best mean test AUC of $\sim 0.89 \pm 0.02$ for a 97\% matching resolution and no blinding.

\begin{figure}[t!]
\centering
\includegraphics[width=\columnwidth]{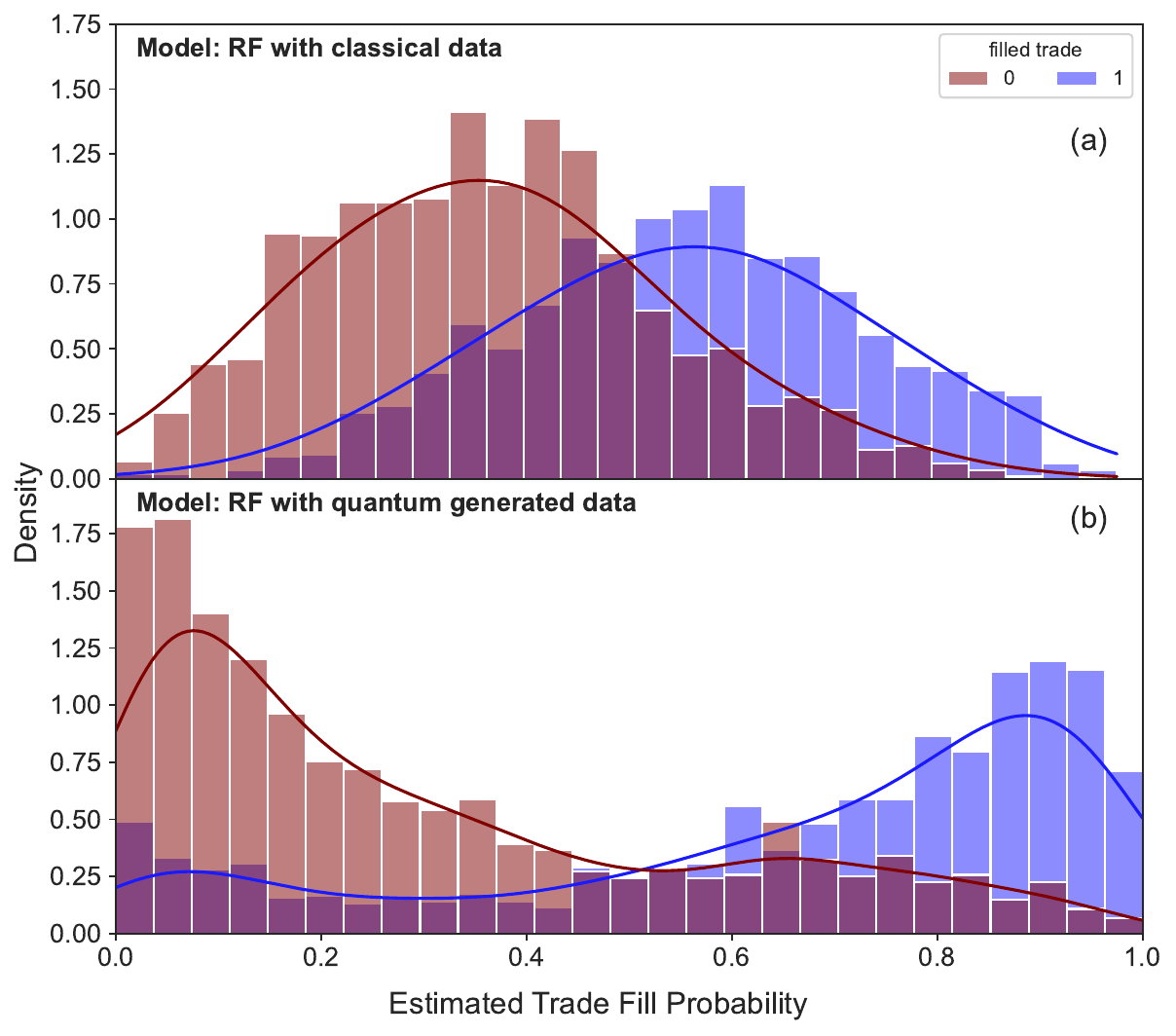}
\caption{Example out-of-sample fill probability distributions from trade backtesting the RF model with 90\% classical-quantum event matching resolution and up to 2 days market blinding, using PQFM features generated by the "longer" circuit on the \texttt{ibm\_torino} device. The predictions of the RF model with classical inputs (a) are compared with quantum-generated inputs (b) with respect to the actual realization of trades filled ("1") or not ("0").}
\label{fig_batch12_RF_prob_dists_80perc10bins_classical_event_matched_qhw_marketblinding2d}
\end{figure}

Finally, Figure~\ref{fig_batch12_RF_prob_dists_80perc10bins_classical_event_matched_qhw_marketblinding2d} shows an example of the predicted fill probability distributions with respect to the actual realized fill results of out-of-sample event-matched trade instances. While the model with classical data has strongly mixing distributions, the model with quantum-generated data more clearly separates the classes correctly.

\section{Conclusion}
\label{sec_conclusion}
The optimization of algorithmic trading strategies is central to the quantitative investment industry and plays at the forefront of financial and economic research that fosters the creation of early test beds for new ideas and emerging technologies, such as quantum computing explored in this work. The domain involves a manifold of statistical models that are limited by the inherent complexity of financial market data and the predictability of associated observables to derive triggers for trading decisions. Among them is the central problem of estimating the fill probability of trade orders with statistical learning algorithms that deal with complex multivariate financial time series with stochastic properties and unknown temporal regularities of varying importance. In this paper, we focus on the execution likelihood estimation for responses to RFQs in the European corporate bond market and empirically investigate, for a select set of machine learning models, the effect of changing their input datasets with quantum-generated features using a particular type of quantum circuit executed on a quantum computer and a noiseless quantum simulator for comparison.

We first introduce a general analysis framework that separates the fill probability estimation process into an online and an offline component. This enables the decoupled offline exploration of slower but more advanced trading data transforms that change the distributions for different learnable pattern associations, which can then be matched to actual online RFQ instances for testing different fill probability estimators. The approach is further supplemented with a classical-to-quantum event matching technique that allows market state-dependent reuse of quantum-generated features and stricter state outcome or label independence controls.

Based on this framework and a given confined market period with real bond trading data and a representative sample of intraday trade instances over 3~months, quantum transformations of respective classical trade event data are used as inputs to four machine learning models and backtested against the same models with the originating classical inputs only. The model performance results for respectively generated fill probability predictions, compared to actually realized trade outcomes, show significant uplifts of out-of-sample test scores for models with noisy quantum hardware-generated inputs over those with noiseless quantum simulation or classical inputs. 

While this is a purely empirical observation of applying a heuristic method to a specific dataset without any generalization guarantees to other market environments or trading datasets, the results are still subject to many open questions. For instance, it is not understood how exactly quantum hardware noise affects our particular quantum circuit, and it is unclear how resulting noise-encoded feature vectors may benefit the analysis of noisy financial observables. This requires further investigation.

Nevertheless, our work in this paper demonstrates how quantum computers, despite their emerging nature today, start to become complementary explorative tools in modeling financial markets and raise questions that may inspire different research directions towards practical applications with error-corrected quantum computers in the future.

\section*{Acknowledgments}
We would like to thank and acknowledge Jae-Eun Park for his leadership in advocating the quantum approach we used in this work and shaping the team's efforts in quantum machine learning. We also thank additional team members whose contributions supported the broader project but were not part of the original draft or core results and finalized approaches presented here. In particular, we thank Vivek Dixet for exploring alternative models and modeling approaches, including different simulated quantum feature maps, classical methods, and meta-modeling approaches. We also thank Andres Ruiz for his extensive work analyzing quantum hardware effects, modeling noise, and developing and testing noisy simulation approaches, and Takahiro Yamamoto for additional contributions to analyzing hardware effects and noise models, as well as Wiktor Mazin for his principal component analyses of classical with respect to different quantum hardware-generated features.

\section*{Author contributions}
This paper was written as part of a collaboration between IBM and HSBC. The authors are listed alphabetically. All authors contributed to project discussions and have reviewed the final manuscript. Additional contributions by co-authors are summarized as follows: D.R. initiated the investigation and established the collaboration on algorithmic trading. J.F., A.C., A. Ciceri, P.I., and D.R. proposed exploring quantum computing for enhancing fill probability estimation in credit algorithmic trading. P.I., M.P., D.R., and A.M. managed the overall project team collaboration and guided the investigation, and B.Q. coordinated the technical team. J.F., A.C., and A. Ciceri provided real-world bond trading insights, the business problem, and expertise to frame it for benchmarking. A.C. prepared the source trading dataset, followed by M.P. to enhance it with higher-dimensional market event representations, as used in this work. B.Q. guided the technical directions and modeling strategy; provided reference implementations; performed selected computations and analyses; and contributed to writing the original draft, final manuscript, and editing. N.S. led the technical development, including modeling pipelines; implemented and evaluated modeling approaches and analyses; conducted experiments with quantum circuits using simulation and hardware; and contributed to the original draft. D.F., K.O., H.H., and K.Y. contributed to the modeling pipelines; explored and evaluated a variety of quantum circuits, modeling approaches, and supportive analyses; and contributed to the original draft. H.K. conducted the primary quantum hardware experiments, performed related analyses, and contributed to the original draft. C.L. explored and evaluated modeling approaches, performed analyses, and contributed to the final manuscript and editing. D.P. guided and performed select validation and result analyses, and contributed to the final manuscript and editing. M.P. led the overall business problem solutioning; developed the framework and methodology for algorithmic trading implementations; invented the classical-quantum event matching technique; managed the trade execution backtesting and benchmarking; designed and analyzed quantum hardware noise experiments; and served as lead author for this manuscript and editorial process.

\begin{table*}[t]
    \renewcommand\thetable{A.3}
    \centering
    \begin{tabular}{llr} \hline
        Model & Hyperparameter & Search Space \\ \hline
        LR & \texttt{solver}
           & ["lbfgs", "sag", "saga", "newton-cholesky", "newton-cg", "liblinear"] \\
           & \texttt{C} 
           & [$10^{-4}$, $10^{-3.2}$, $10^{-2.4}$, $10^{-1.6}$, $10^{-0.8}$, $10^{0}$, $10^{0.8}$, $10^{1.6}$, $10^{2.4}$, $10^{3.2}$, $10^{4}$] \\
           & \texttt{penalty} 
           & "l2" \\
           & \texttt{max\_iter} & 10000 \\

        XGB & \texttt{max\_depth} 
            & [3, 5, 7, 9, 11] \\
            & \texttt{n\_estimators}
            & [80, 100, 120, 140, 160, 180] \\
            & \texttt{learning\_rate} 
            & [0.15, 0.1, 0.05, 0.01] \\
        
        RF & \texttt{criterion} 
           & ["gini", "entropy", "log\_loss"] \\
           & \texttt{n\_estimators} 
           & [80, 100, 120, 140, 160, 180] \\
        
        NN & \texttt{hidden\_layer\_sizes}
           & [(3), (100), (164), (164, 65), (262, 164, 98), (262, 196, 131, 65), \\
           & &  (262, 196, 131, 65, 3), (196, 3)] \\
           & \texttt{activation} 
           & ["relu", "logistic", "tanh"] \\
           & \texttt{learning\_rate\_init} 
           & ["constant", "invscaling", "adaptive"] \\
           & \texttt{max\_iter}
           & 10000 \\ \hline
        \end{tabular}
    \caption{Search space for tuning selected hyperparameters of models used in this paper, as defined in the API documentation of \texttt{scikit-learn}~\cite{scikitlearn}. Note: the hidden layer sizes for the NN model represent only one example, as they are dynamically generated for given data sample dimensions.}
    \label{tab_model_params_grid_OOS}
\end{table*}
%
\section*{Declaration of competing interest}

The authors declare that they have no known competing financial interests or personal relationships that could have appeared to influence the work reported in this paper.

\section*{Disclaimers}
This paper was prepared for information purposes and is not a product of HSBC Bank Plc. or its affiliates. Neither HSBC Bank Plc. nor any of its affiliates make any explicit or implied representation or warranty and none of them accept any liability in connection with this paper, including, but not limited to, the completeness, accuracy, reliability of information contained herein and the potential legal, compliance, tax or accounting effects thereof.\\
\\
\copyright~IBM \& HSBC Group 2025
\newpage
%
\appendix

\section{Technical settings for backtesting}\label{app_1}

\begin{table}[b!]
    \centering
    \begin{tabular}{ll} \hline
        Backtesting Parameter & Setting \\ \hline
        dataset parameter ($\eta$)
        & classical (Section~\ref{subsec_data}) \\
        & simulation (Section~\ref{subsubsec_qsim}) \\
        & hardware "short" (Section~\ref{subsubsec_qhw}) \\
        & hardware "long" (Section~\ref{subsubsec_qhw}) \\
        & event matching (Section~\ref{subsubsec_cqem}) \\
        model ($m$)
        & Table~\ref{tab_model_params_grid_OOS} \\
        training size ($\mathcal{T}$)
        & 500 \\
        & 1000 \\
        & 1500 \\
        & 2000 \\
        blinding window ($\Delta_{ka}$)
        & $\geq 1~\mu s$ to $< 5$ trading days \\
        \hline
    \end{tabular}
    \caption{Backtesting runtime settings for the results presented in this paper.}
    \label{tab_backtesting_params}
\end{table}
The testing setup uses the Python programming language with various open-source packages. The machine learning models Logistic Regression (LR), Random Forest (RF) and Feed-Forward (also referred to as multi-layer perceptron) Neural Network (NN) use the implementation in \texttt{scikit-learn}~\cite{scikitlearn}, while for Gradient Boosting we use the implementation of eXtreme Gradient Boosting (XGB)~\cite{chen2016xgboost} as a representative of modern variants of gradient boosting, which are still among the state-of-the-art methods for tabular data~\cite{grinsztajn2022why, shwartz2022tabular, ye2025closerlookdeeplearning}, as found in this work. The hyperparameters of these models are extensively tuned to optimize model performance using the \texttt{GridSearchCV} module of \texttt{scikit-learn} and the parameter grids specified in Table~\ref{tab_model_params_grid_OOS}. For parameters that are not tuned, we use the default parameters of the respective model implementation.

\vspace{0.5em}
The final configuration of the backtesting protocol as defined in Section~\ref{subsec_backtesting} and used to produce the results reported in Section~\ref{subsec_resbacktesting} and \ref{subsec_reseventmatching} is shown in Table~\ref{tab_backtesting_params}.

\bibliographystyle{elsarticle-num} 

\end{document}